\documentclass[amsmath,11pt,amssymb]{revtex4}

\usepackage{multibbl}
\usepackage{color}
\usepackage[]{graphicx}
\usepackage{latexsym}
\usepackage{natbib}
\usepackage{amsmath}
\usepackage[justification=raggedright,font=small,labelfont=bf]{caption}
\usepackage[caption=false]{subfig}
\usepackage{multirow}
\usepackage{mathtools}
\usepackage{textcomp}

\usepackage[pdftex,hyperfigures,breaklinks,colorlinks,citecolor=black,linkcolor=black]{hyperref} 

\newcommand{\be}{\begin{equation}}
\newcommand{\ee}{\end{equation}}
\newcommand{\bc}{\begin{cases}}
\newcommand{\ec}{\end{cases}}
\newcommand{\core}{\rm{core}}

\renewcommand{\figurename}{Fig.}

\usepackage[final]{feynmp}
\begin{document}

\title{The k-core as a predictor of structural collapse in mutualistic ecosystems}






\author{Flaviano Morone, Gino Del Ferraro, and Hern\'an A. Makse}

\affiliation{Levich Institute and Physics Department, City College of
  New York, New York, NY 10031}




\begin{abstract}
\begin{center}
\textbf{\abstractname}
\end{center}
{\bf Collapses of dynamical systems into irrecoverable states are
  observed in ecosystems, human societies, financial systems and
  network infrastructures. Despite their widespread occurrence and
  impact, these events remain largely unpredictable. In searching for
  the causes for collapse and instability, theoretical investigations
  have been so far unable to determine quantitatively the influence of
  the structural features of the network formed by the interacting
  species. Here, we derive the condition for the stability of a
  mutualistic ecosystem as a constraint on the strength of the
  dynamical interactions between species and a topological invariant
  of the network: the k-core. Our solution predicts that when species
  located at the maximum k-core of the network go extinct as a
  consequence of sufficiently weak interaction strengths the system
  will reach the tipping point of its collapse. As a key variable
  involved in collapse phenomena, monitoring the k-core of the network
  may prove a powerful method to anticipate catastrophic events in the
  vast context that stretches from ecological and biological networks
  to finance.}
\end{abstract}

\maketitle

\section{Introduction}

A complex dynamical system collapses when a small perturbation 
in the parameters characterizing the species interactions causes a 
large-scale extinction of the species in the system
\cite{may3,strogatz,caldarelli,shlomo,sheffer1,sheffer2,allesina,gao,bascompte-linear,coyte,may-banking, guido-banking}.
The tipping point at which the system suddenly shifts to the
irrecoverable state is, for practical purposes, the most important
quantity one wishes to know~\cite{sheffer1,sheffer2,gore}. It is a
function of the dynamical and structural parameters of the system
determined by the fixed point solution of the nonlinear equations
describing the system's dynamics~\cite{may3}.  However, the tipping
point is hard to determine, due to the difficulties encountered in
solving the nonlinear dynamical equations to quantify the dependence
of the fixed point solution on the system parameters and, in
particular, on the features of the underlying network of interacting
species in the system~\cite{may3, caldarelli,gao,sheffer2}. Indeed, no
exact analytical result exists, so far, that relates the network
properties to the fixed points of the dynamical system.  Here, we
first study numerically the fixed point equations of a dynamical
system of mutualistic species and then derive the analytical solution
to compute the tipping point using a logic approximation. Our solution
reveals that the root cause of the system collapse is the extinction
of species located in the maximum k-core of the network.


The concept of k-core was introduced in social sciences
~\cite{seidman} to define network cohesion and was then applied in
many other contexts \cite{wormald}, including the robustness of random
networks~\cite{dorogotsev}, the structure of the internet~\cite{carmi,
  hamelin}, viral spreading in social networks~\cite{gallos}, the
large-scale structure of brain networks~\cite{sporn}, and the jamming
transition \cite{kate}. For a network of interacting species, the
k-core is the portion of the network that remains after iteratively
removing from the network all species linked to fewer than $k$ other
species (see Figs.~\ref{fig:fig1}a, b and Supplementary Information Section
~\ref{sec:kcore} ~\cite{seidman, dorogotsev}).  For a given $k$, the
subset of species in the k-core consists of the periphery, called the
k-shell, and the remaining k+1-core; therefore, the k-shell is the
region of the k-core not included in the k+1-core
(Fig.~\ref{fig:fig1}a).  Thus, the network has a nested structure of
k-cores with increasing k-shells of order $k_s$, starting from the
periphery of the network or 1-shell, $k_s=1,$ and its 1-core which
includes all the network (except for isolated nodes). The 1-core
contains the 2-core, and so on, up to the innermost core of the
network which is the maximum k-core labeled by the ``k-core number''
$k^{\rm max}_{\rm core}$. The k-core number is a topological invariant
of the network \cite{dorogotsev}.

\section{Model of a mutualistic ecosystem}

 We consider complex systems populated by $N$ interacting species,
 also referred to as network nodes, whose directed interactions can be
 graphically portrayed as links in a network via the adjacency matrix
 $A_{ij}$ such that $A_{ij}=1$ if species $i$ interact with $j$, and $A_{ij}=0$
 otherwise. In general $A_{ij}\neq A_{ji}$ for directed networks.
The strength of the directed interaction from species $i$ to $j$ is
$\gamma_{ij}$. In this paper we consider the case of mutualistic
ecosystems where organisms of different species cooperate with each
other by benefiting from the activities of the other, such as plants
and pollinators. These systems are characterized by positive
interactions between the species, $\gamma_{ij}>0$.
Dynamical systems with positive and negative interactions, such as
neuron, genes or predator-prey ecosystems, are out of the scope of the present work
and will be treated in a follow-up.

The state of the system is encoded in the multiplet of species
densities $\vec{x}(t)\equiv(x_1(t), \dots, x_N(t))$ evolving in time
towards a fixed point $\vec{x}^*$, where $\dot{\vec{x}}^*=\vec{0}$
~\cite{strogatz}.  When the species do not interact, 
i.e. for $\gamma_{ij}=0$, each species density changes through time as
$\dot{x}_i(t)=f_i(x_i)$, and the fixed points are found by solving
$f_i(x_i^*)=0$ for all $i$. When the species interact according to
$\gamma_{ij}$, $x_i(t)$ is influenced by the densities $x_j(t)$ of the
species linked to it in the network of interactions. While these
interactions are generally complex,
it is generally recognized that they saturate when the density of
interacting species increases~\cite{may,holland,alon,amit,thebault,
  bastolla}.
This occurs in mutualistic interactions between species in ecosystems,
for which the benefit accorded by one species to another saturates to
a limiting value~\cite{may,holland,thebault,bastolla}.
In biology, the expression level of gene products are modeled by Hill
or sigmoidal response functions which saturate at high concentrations
of the interacting gene (SI Section~\ref{gene})
~\cite{alon}. Enzymatic reactions are also modeled by Hill functions
in the Michaelis-Menten equation ~\cite{alon} and firing rates of
neurons saturate at high membrane potential via sigmoidal
functions~\cite{amit, sompolinsky}.

In the following, we treat explicitly the paradigmatic case of
dynamical systems of ecological mutualistic networks, but the results
we obtain hold true for the larger class of nonlinear systems where a
Hill or sigmoidal function models the interactions.
A network of mutualistic species describes a system of symbionts
obligated to each other because they cannot survive
independently~\cite{thebault, holland, may}, e.g., an ecosystem of
plants and pollinators (Fig.~\ref{fig:fig1}b).  The dynamics of
species densities, $x_i(t)$, interacting via the network $A_{ij}$ with
directed and positive interaction strengths $\gamma_{ij}$, is
described by the following set of nonlinear differential
equations~\cite{holland,thebault,may,bastolla,holland3}:

\begin{equation}
\dot{x_i}(t)\ =\ -d x_i - s x_i^2 + \sum_{j=1}^{N}A_{ij}\gamma_{ij}
\frac{x_ix_j}{\alpha+\sum_{k=1}^{N}A_{ik}x_k}\ , \,\, i\in\{1, \cdot
\cdot \cdot, N\}\ .
\label{eq:model}
\end{equation} 
Here $d>0$ is the death rate of the species, $s>0$ is the
self-limitation parameter modelling the intraspecific competition that
limits a species' growth once $x_i$ exceeds a certain value, $\alpha$
is the half-saturation constant, and $\gamma_{ij}>0$ is the
mutualistic interaction strength between species $i$ and $j$
characterizing the strength of the nonlinear interaction term.  The
dynamical parameters $(\{\gamma_{ij}\}, d, s, \alpha)$ have been
extensively discussed in the literature
~\cite{holland,thebault,may,bastolla,holland3}.  The network is
bipartite between, e.g., plants and pollinators
(Fig.~\ref{fig:fig1}b).
Our goal is to bridge the gap from structure to dynamics by obtaining
the fixed point solution of dynamical equations to predict the
tipping point of collapse in terms of a feature of the network.


\section{Numerical analysis of the ecosystem collapse}

We start by performing a numerical study of the tipping point of the
system (details in Methods Section and in SI Section \ref{sec:phase_diagram}), 
and then we elaborate our analytical solution based on 
approximations supported by the numerical evidence. 
Figures~\ref{fig:fig2}b-f show the numerical solution of the
Eqs.~\eqref{eq:model} for different parameters on a real
plant-pollinator mutualistic network from the Chilean Andes obtained
from Ref.~\cite{arroyo} (Net \#10 in Supplementary
Table~\ref{table:mutualistic_net}).  We plot the fixed point average
density (properly rescaled) $\langle x^*\rangle=N^{-1}\sum_ix_i^*$
as a function of $K_{\gamma}=\frac{\alpha s(\gamma+d)}{(\gamma-d)^2}$,
which is the main control parameter that determines the collapse of
the system according to the theoretical solution
Eqs. (\ref{eq:fixpoint_canonical}).  Here $\gamma$ is the average
interaction strength and $K_{\gamma} \approx 1/\gamma$ since
$d\ll\gamma$.

By increasing $K_{\gamma}$ or, analogously, decreasing the
interactions $\gamma$, we find that for all the numerical ecosystems in
Fig. \ref{fig:fig2} there exist a point of collapse at a given critical value $K_{\gamma_c}$
(or analogously $\gamma_c$), which is the tipping point of the
ecosystem.  This collapse is exemplified by the transition from a
non-zero fixed point $\langle x^*\rangle \neq 0$ for $K_{\gamma} <
K_{\gamma_c}$ where the species are alive to a zero fixed point
$\vec{x}^*=\vec{0}$ for $K_{\gamma} > K_{\gamma_c}$ that corresponds
to the extinction of all species
~\cite{thebault,holland,holland3,gao,may2}.

The collapsed phase corresponds to the trivial fixed point of
Eqs.~\eqref{eq:model}, $\vec{x}^*=\vec{0}$.
The decrease of the interaction $\gamma$ that drives the system to
collapse for $\gamma<\gamma_c$ could be caused, for example, by
external global conditions such as changes in environmental conditions
like global climate change. These global changes produce shifts in
phenology and hence changes in the interaction strength $\gamma$ that
affect all species~\cite{sheffer1,sheffer2}. The question is then how
to predict this tipping point.




\section{Analytical solution of the ecosystem collapse}

We first show that the fixed point equations for this system can be
written in terms of the Hill function \cite{alon,holland}.  We
consider a system with $\gamma_{ij}=\gamma$ (see Methods) and make a
change of variables to the reduced density:
\begin{equation}\label{eq:y}
y_i^*=\frac{s}{\gamma-d}\sum_{j=1}^NA_{ij}x^*_j,
\end{equation}
whereby the fixed point equations can be written using a sum of Hill
functions of the form $H_1(x,T) = x/(T+x)$, where $T>0$ is the
half-saturation constant \cite{alon,holland} (details in SI Section ~\ref{sec:undirected_net}):
\begin{equation}
y_i^*\ = \ \sum_{j=1}^NA_{ij}
H_1\Bigg(y_j^*-\frac{\alpha ds}{(\gamma-d)^2},\ \frac{\alpha\gamma s }{(\gamma-d)^2}\Bigg)\ .
\label{eq:fixpoint_y}
\end{equation}
The Hill function $H_1$ is the first of a family of response functions
parametrized by the Hill coefficient $n$ as $H_n(x,T)= x^n/(T^n+x^n)$,
where $n$ characterizes the degree of cooperativity among the
interacting species~\cite{alon,alon1}.  This particular
interaction term in Eqs. (\ref{eq:model}) is not crucial for the
solution of the problem: any saturating sigmoidal-like function will
lead to the k-core collapse of the dynamical system (SI Section
\ref{gene}).

A widely used approximation to treat these systems analytically
involves the logic approximation of the Hill function as proposed by
Kauffman \cite{kauffman2} to describe genetic Boolean networks
\cite{kauffman}.  This approximation assumes $n\to \infty$ and
replaces the interaction function by a logic ON and OFF switch
according to whether the input $x$ is above the threshold $T$ or
below, respectively. That is, it replaces $H_1$ by $H_1(x,
T)\approx\Theta(x-T)$, where the Heaviside function $\Theta(x) = 1$ if
$x>0$ and zero otherwise. Both, the continuous description for finite
$n$ and its Boolean-logic approximation for $n\to \infty$ are also
widely used to describe artificial and real neural networks
\cite{amit}.  Inspired by these works \cite{alon,amit,
  kauffman,kauffman2}, we apply the logic approximation to
Eqs. ~\eqref{eq:fixpoint_y} to solve the model analytically (SI Section
\ref{sec:limits} systematically investigates numerically the limit of
validity of the logic approximation).

By using the logic approximation of the Hill function~\cite{alon,amit,
  kauffman,kauffman2}, i.e. $H(x,T)\to\Theta(x-T)$, the fixed point
equations can be recast in the following analytically tractable form:
\begin{equation}
\begin{aligned}
y_i^*\ &=\ \sum_{j=1}^NA_{ij}\Theta(y_j^*-K_{\gamma})\ ,\\ K_{\gamma}\ &=\ \frac{\alpha
  s(\gamma+d)}{(\gamma-d)^2}\ ,
\end{aligned}
\label{eq:fixpoint_canonical}
\end{equation}
where 
$K_{\gamma}$ is the threshold on the mutualistic benefit;
the subscript emphasizes its main dependence on $\gamma$, $K_{\gamma}
\propto 1/\gamma$ (Fig. \ref{fig:fig4}a) since $d \ll \gamma$.
Concretely, $K_{\gamma}$ is the threshold of the $\Theta$-function in
Eqs. ~\eqref{eq:fixpoint_canonical}, which allows species $i$ to
benefit from mutualistic interactions with species $j$ only when their
densities $y_j^*$ are bigger than $K_{\gamma}$. Weak interactions
$\gamma$ correspond to large thresholds $K_{\gamma}$, which, by
inhibiting the benefits $y_j^*$ conferred by species $j$ to $i$,
produce small values of $y_i^*$.  Thus, if $\gamma$ falls below the
critical value $\gamma_c$, no mutualistic benefit is exchanged among
species since the corresponding critical threshold,
\begin{equation}
  K_{\gamma_c} = \frac{\alpha s(\gamma_c+d)}{(\gamma_c-d)^2},
\label{kgammac}
\end{equation}
is too high, and the entire system collapses via a catastrophic
transition to the state $\vec{x}^*=\vec{0}$ (Fig.~\ref{fig:fig4}a), as
shown numerically in Figs. \ref{fig:fig2}b-f.

Next we show that the critical interaction strength for collapse,
$\gamma_c$ (or $K_{\gamma_c}$) is determined by the maximum k-core of
the network~$k^{\rm max}_{\rm core}$.
The reduced density $y_i^*$ assumes only integer values in the set
$y_i^*\in\{1,\dots,k_i\}$, where $k_i$ is the degree of, or number of 
species interacting with, species $i$. 
Therefore, to solve for $y_i^*$ at a given threshold $K_\gamma$, we
remove all species $j$ with degree $k_j<K_{\gamma}$ from
Eqs. ~\eqref{eq:fixpoint_canonical}, since these species give zero
contribution to the right hand side of
Eqs. ~\eqref{eq:fixpoint_canonical}, and we only solve for the
remaining species. The procedure is graphically explained in
Fig.~\ref{fig:fig4}b for a simple ecosystem with a maximum 2-core and
interaction strength that could be anywhere between $1<K_\gamma<2$,
and in SI Section~\ref{sec:small_net}, for fully
connected networks of 2, 3, and 4 species.

After these first removals are done (Step 1 in
Fig.~\ref{fig:fig4}b), the species left in the network have
smaller degree $k_j'$, and we perform a new wave of removals of
species $j'$ if $k_j'<K_{\gamma}$ (Step 2 in
Fig.~\ref{fig:fig4}b). This pruning process stops when the
degree of each remaining species is larger than $K_{\gamma}$. The
process we just described is precisely the algorithm for extracting
the $K_{\gamma}$-core of the network~\cite{seidman,dorogotsev,gallos},
as explained in Fig.~\ref{fig:fig1}b:
iteratively removing all species from the network with degree $k< \lceil
K_{\gamma}\rceil$, where $\lceil \cdot \cdot \rceil$ denotes the 
ceiling function.  Thus, the nodes remaining at the end of this pruning
process, if any, form a $K_{\gamma}$-core by construction, as 
shown in the Step 3 of Fig.~\ref{fig:fig4}b.
Since $y_i^*$ in Eqs.~\eqref{eq:fixpoint_canonical} measures the 
number of links of species $i$ to this remaining $K_{\gamma}$-core, we find the 
nontrivial fixed point solution for the species belonging to this 
$K_\gamma$-core as (Fig.~\ref{fig:fig4}b Step 3):
\begin{equation}
y_i^*\ =\ {\rm number\ of\ links\ of\ species\ }i{\rm \ to \ species\ in\ the \ }
  K_{\gamma}{\rm-core}\ \equiv\ \mathcal{N}_i(K_{\gamma}) .
\label{eq:y_solution}
\end{equation}
Equation~\eqref{eq:y_solution} remains valid also for the species 
placed outside the $K_{\gamma}$-core, since the only nonzero benefits 
they receive come from species inside the $K_{\gamma}$-core 
(Fig.~\ref{fig:fig4}c). 
Indeed, for a species $i$ outside the 
$K_{\gamma}$-core, $y_i^*$ may be nonzero only if species $i$ interacts with 
at least one species inside the $K_{\gamma}$-core. However, those 
species for which $0<y_i^*< \lceil K_{\gamma}\rceil$ have no influence in the
ecosystem, meaning that their disappearance does not change the
density of any other species. In practice, they are commensalists rather 
than symbionts, that benefit from the species located in the 
$K_{\gamma}$-core without benefiting nor damaging them, as seen 
in Fig.~\ref{fig:fig4}c.

\section{Tipping point predicted by the maximum k-core of the network}

Equations~\eqref{eq:y_solution} reveals how the dynamics is intertwined
with the network structure through the number of links to the
$K_{\gamma}$-core, $\mathcal{N}_i(K_{\gamma})$.
Indeed, when these links disappear, the system collapses. Since the
densities must be positive by definition, $x_i^*>0$ ($x_i^*$ is
obtained from Eqs.~\eqref{eq:y_solution} by a change of variables,
see Supplementary Eqs.~\eqref{eq:x_solution}), hence $y_i^*$ must
also be positive. Then, we must have
$\mathcal{N}_i(K_{\gamma})>0$. However, this condition cannot be
satisfied by any species $i$ if $K_{\gamma}>k_{\rm core}^{\rm max}$,
because when the threshold $K_{\gamma}$ in
Eqs.~\eqref{eq:fixpoint_canonical} is larger than the maximum
k-core of the network, the number of links to the $K_\gamma$-core is,
by definition, zero, i.e., $\mathcal{N}_i(K_{\gamma}>k_{\rm core}^{\rm
  max})=0$.  As a consequence, if $\gamma$ is reduced to the point
that $K_{\gamma}$ is slightly above the maximum k-core $k_{\rm
  core}^{\rm max}$, so that $K_{\gamma} > k_{\rm core}^{\rm max}$, the
system collapses to the state $x_i^*=0$, where the species are
extinct.
The critical value $\gamma_c$ at this tipping point of collapse is
predicted as:
\begin{equation}
K_{\gamma_c}\ =\  k_{\rm core}^{\rm max}\ \ \to\ \ 
\alpha s\frac{(\gamma_c+d)}{(\gamma_c-d)^2}\ =\  k_{\rm core}^{\rm max}\ ,
\label{eq:critical_T}
\end{equation}
which represents our main result
relating the dynamical parameters at the tipping point 
with a global topological network property. 

We confirm the main theoretical result of Eq. \eqref{eq:critical_T}
with a numerical simulation using the same mutualistic network of
Fig. \ref{fig:fig2} (Net \#10 in Supplementary
Table~\ref{table:mutualistic_net}, Fig. \ref{fig:fig5}a shows its k-shell
structure). Figure \ref{fig:fig5}b shows the fixed point average
density $\langle x^* \rangle$ for this system which confirms that the
collapse of the ecosystem occurs when $K_{\gamma_c}$ satisfies the
critical condition~\eqref{eq:critical_T}. That is, the system
collapses at $K_{\gamma_c}=4$ which corresponds to the maximum k-core
for this network, $k_{\rm core}^{\rm max}=4$ (Fig. \ref{fig:fig5}a).
We also compare the logic approximation (black curve) to the numerical
solutions in Figs. \ref{fig:fig2}b-f and Fig. \ref{fig:fig2}g, which
corresponds to the case where $\Delta=0$.  Figure \ref{fig:fig2}h
plots the numerical tipping point $K_{\gamma_c}$ compared to the k-core
prediction for this network $k_{\rm core}^{\rm max}=4$. We find that
the logic approximation captures well the tipping point of the system
across realistic values of death rates parameters $d \in
[0.1-0.3]$\cite{thebault,holland} (further examinations are provided
in SI Section \ref{sec:limits}).


As the interaction strength decreases (so $K_\gamma$ increases) 
due to external global conditions, the system suffers a series of 
partial collapses characterized by the sharp drops in the species 
density as shown in Fig.~\ref{fig:fig5}b, at precise integer values of $K_\gamma$ 
equal to the index $k_s$ of each k-shell, in succession. This occurs
until the species in the maximum k-core at $k_{\rm core}^{\rm max}=4$
go extinct with the concomitant collapse of the entire network.
Therefore, as the strength of mutualistic interaction $\gamma$
decreases, the species in the outer k-shells (the ``leaves'' in the
network) go extinct first, while species in the innermost k-core
survive up to the tipping point of total collapse (insets in
Fig.~\ref{fig:fig5}b). As a consequence, Eq. ~\eqref{eq:critical_T}
can be used as a warning signal for the proximity of the system to the
tipping point by measuring independently the dynamical parameters and
the k-core number of the network.

In SI Section \ref{metrics} we compare the prediction of the tipping point of the system made by the $k_{\rm core}^{\rm max}$ to the prediction of collapse made by other metrics, such as nestedness \cite{bascompte,olesen}, spectral radius \cite{staniczenko}, and connectance (average degree). Supplementary Fig. 4 shows the result of this comparison. Overall, the metrics which are related to the $k_{\rm core}^{\rm max}$ via mathematical bounds, e.g. the spectral radius ($\rho \geq k_{\rm core}^{\rm max}$ \cite{bickle}) and the connectance, are also good predictors of the tipping point when these bounds are saturated. In more general conditions, though, i.e. far from saturated bounds, the $k_{\rm core}^{\rm max}$ remains the metric which theoretically predicts the collapse of the system.

\section{Stability analysis and phase diagram of system feasibility}

Once we have the solution~\eqref{eq:y_solution} to the fixed point
equations, we can study its local stability, which is controlled by
the stability matrix $\mathcal{M}_{ij}=\frac{\partial
  \dot{x_i}}{\partial x_j}\Big|_{\vec{x}^*}$. Indeed, stability theory
has been at the core of the ecosystem literature since May \cite{may3}
posed the question whether a large ecosystem would become stable or
unstable, see below.  

A fixed point solution~\eqref{eq:y_solution} is locally stable if all
the eigenvalues $\lambda^{\mathcal{M}}_i$ of the stability matrix
$\hat{\mathcal{M}}$ have negative real part \cite{may3}.
These eigenvalues can be calculated analytically in our model. We find
(SI Section~\ref{sec:linear_model}):
\begin{equation}
\lambda^{\mathcal{M}}_i\ =\ -\gamma \frac{\mathcal{N}_i(K_\gamma)}
{K_\gamma + \mathcal{N}_i(K_\gamma)}\ ,\ \ \ \ 
i = 1,\dots, N\ ,
\label{eq:eigen_stability_matrix}
\end{equation}
which are, in fact, all negative. Therefore our solution, if it exists,
is always stable. This result has important consequences as we show
next.

Interestingly, the largest and thus most critical eigenvalue
$\lambda^\mathcal{M}_{\rm max}$ is the one corresponding to the
commensalist species $i$ with the minimal number of links
$\mathcal{N}_i$ to the symbionts located in the $K_\gamma$-core
(Fig. \ref{fig:fig4}c). The most critical species, i.e., species most
exposed to extinction, are commensalists with a single link to the
$K_\gamma$-core with $\lambda^\mathcal{M}_{\rm max} =
-\gamma/(K_\gamma+1)$. As the system approaches its collapse, the
commensalists with the fewest number of links to the $K_\gamma$-core
go extinct first. Such a dynamics is clearly seen in the sketches of
Fig.~\ref{fig:fig4}a and the network panels of the numerical solution
in Fig.~\ref{fig:fig5}b.

Thus, our solution predicts that the system's approach to the tipping
point of collapse is signaled by an increase of commensalist species
at the outer shells, and a reduction of symbionts at the inner cores.
From Eqs. ~\eqref{eq:eigen_stability_matrix} we also conclude that
when $K_\gamma > k_{\rm core}^{\rm max}$, all the eigenvalues vanish,
thus the feasible fixed point becomes unstable (and also unfeasible),
with the concomitant extinction of all species. This confirms the
tipping point Eq.~\eqref{eq:critical_T} derived above from the
existence of the feasible nontrivial solution.

These considerations lead to the phase diagram of feasible and stable
mutualistic ecosystems depicted in Fig.~\ref{fig:fig5cd}a
in the space $(K_{\gamma}, k_{\rm
  core}^{\rm max})$.
The phase diagram features the predicted `tipping line' of instability
defined by the condition~\eqref{eq:critical_T}, which separates the
feasible-stable phase:
\begin{equation}
 K_{\gamma} < k_{\rm core}^{\rm max} \ \ \ \ \ 
\mbox{(condition of existence of the feasible-stable state)}\ ,
\label{eq:stable}
\end{equation}
from the collapsed phase:
\begin{equation}
K_{\gamma} >   k_{\rm core}^{\rm max }\ \ \ \ \ \mbox{(condition of
    collapsed state)}\ .
\end{equation}

We test this phase diagram by plotting the values of $(K_{\gamma},
k_{\rm core}^{\rm max })$ obtained from real mutualistic networks of
plant-pollinator and plant-seed dispersers
~\cite{bascompte-linear,thebault} (see Fig. ~\ref{fig:fig5cd}a, Supplementary Table
~\ref{table:mutualistic_net}, and SI
Section~\ref{sec:plant_pollinator}). All real mutualistic ecosystems
lie in the feasible-stable region situated above the tipping line, in
agreement with the theory.

This conclusion contrasts with the prediction obtained by
approximative linear stability methods introduced by May
in~\cite{may3} based on Wigner's semicircle law frequently used in the
literature \cite{allesina,coyte}. This approach considers a linear
model of species interactions rather than the sigmoidal Hill function.
The stability matrix is then
$\mathcal{M}'_{ij}\ =\ -\delta_{ij}+A_{ij}/K_\gamma$, and, assuming a
random adjacency matrix $A_{ij}$, it is computed with random matrix
theory \cite{may3,allesina}. The stability condition on the negativity
of the real part of the most critical eigenvalue of
$\mathcal{M}'_{ij}$ is now given by $\lambda^A_{\rm max}< K_\gamma$,
where $\lambda^{A}_{\rm max}$ is the largest eigenvalue of $A_{ij}$
(see SI Section \ref{may_stability} and \cite{may3}).
The condition $\lambda^A_{\rm max}< K_\gamma$ leads to the May's
diversity-stability paradox \cite{may3} by which an ecosystem would
become unstable upon increasing the diversity of the species.  This
prediciton is valid for any type of ecosystem, and in
particular for a mutualistic ecosystem, leading to the
paradoxical result that cooperation destabilizes the ecosystem.
This paradox arises because $\lambda^A_{\rm max}$ increases with the
number of species in the ecosystem \cite{may3} and therefore,
diversity as measured by the number of species, has a destabilizing
effect.

On the contrary, 
our nonlinear theory predicts the opposite result.  First, we predict
that mutualistic interactions are beneficial for the ecosystem: for a
given network structure (fixed $k_{\rm core}^{\rm max}$), systems with
larger $\gamma$ tend to be more robust since condition
Eq. \eqref{eq:stable} is easier to satisfy.
Second, as the diversity of the ecosystem, measured as number of
symbionts in the maximum core $k_{\rm core}^{\rm max}$, increases, the
value $k_{\rm core}^{\rm max}$ increases, hence the condition Eq.
(\ref{eq:stable}) is also easier to satisfy in this case.  Therefore,
diversity of symbionts at the maximum core of the network increases
the stability of the system.

Thus, we show that the analytical solution of the nonlinear model
resolves the long-standing diversity-stability paradox \cite{may3} in
mutualistic ecosystems by introducing a new principle of stability.
This principle states that the more symbionts there are in the maximum
core of the network the higher the robustness. Thus, diversity, mutualism and cooperation 
stabilizes the ecosystem  rather than the opposite as paradoxically proposed in
~\cite{may3}.  
 Our results highlight the importance of considering the
exact stability analysis of the nonlinear model Eqs.
~\eqref{eq:model} instead of the linear model when reaching conclusions about the
stability of ecosystems.  Indeed, studies of the microbiome
\cite{coyte} based on the linear model and Wigner's semicircle law
have concluded that cooperating networks of microbes in the human gut
are often unstable, in contrast to empirical evidence.

\section{ Summary}

We presented an analytic solution of the tipping point for a nonlinear
model of mutualistic dynamical systems in terms of a topological
invariant of the network, the k-core number.
The k-core structure of the network privileges the species at the
inner k-core, which are the ``keystone species''~\cite{microbiome}
like the plant Angelica pubescens in Net \#2, Fig.~\ref{fig:fig5cd}b.
These keystone species are analogous to ``influencers'' in social
networks~\cite{gallos,morone} that guarantee the integrity of the
entire ecosystem. Therefore, species at the innermost core should be
protected first for the sake of the whole ecosystem.

Since our theoretical results are applicable to a large class of
systems governed by nonlinear Hill, logistic or sigmoidal
interactions, the conclusions could be equally applicable to other
complex systems.  Drawing analogies from financial and banking
ecosystems \cite{may-banking,guido-banking}, to neural circuitry
\cite{amit,sompolinsky}, microbial ecosystems~\cite{coyte, biroli}, 
and gene regulatory networks~\cite{alon,kauffman,kauffman2}, our
results provide the way to avoid systemic risks built in these systems
by protecting the system's vital core.

\bigskip
\bigskip
\noindent
{\bf \large Data availability}

Data that support the findings of this study are publicly available at the Interaction Web Database at
\url{https://www.nceas.ucsb.edu/interactionweb/}.

\noindent
{\bf \large Acknowledgments} 
\noindent

Research was sponsored by NSF-IIS 1515022, NIH-NIBIB R01EB022720,
NIH-NCI U54CA137788/U54CA132378, and Army Research Laboratory under
Cooperative Agreement W911NF-09-2-0053 (ARL Network Science CTA). We
are grateful to Soff\'ia Alarc\'on for discussions.

\noindent
{\bf \large Author contributions}

All authors contributed equally to all parts of the study.

\noindent
{\bf \large Additional information}

{\bf Supplementary Information} accompanies this paper at \url{https://www.nature.com/nphys}.

\noindent
{\bf \large Competing interests} 

The authors declare no competing interests. 

\noindent
{\bf Correspondence and requests for materials} should be addressed to H. A. M.




\bigskip


\clearpage 

\section{Methods}

\subsection{Numerical integration}
\label{numerical}

In general, the interaction strengths between species in ecological
networks are weighted and directed \cite{bascompte-linear},
i.e. $\gamma_{ij}\neq\gamma_{ji}$, meaning that the effect of species
$i$ upon species $j$ is different from the effect of species $j$ upon
$i$ and also the interaction strengths are all different
(Fig. \ref{fig:fig2}a).  This heterogeneity is relevant to the
stability of coupled systems and it is important then to study their
relevancy to the determination of the tipping point.  Therefore, we
study the influence of weighted interactions to the location of the
tipping point of the dynamical system given by
Eqs.~\eqref{eq:model}. 

For our numerical investigation, we characterize the weights of the
interactions $\gamma_{ij}$ by a probability distribution with mean
value $\gamma$ and width $\Delta$.  Following \cite{thebault,holland},
we take $\gamma_{ij}$ as i.i.d. random variables drawn from the
uniform distribution $P(\gamma_{ij}) = \frac{1}{2\Delta}$, if
$\gamma-\Delta\leq\gamma_{ij}\leq \gamma+\Delta$, and zero otherwise
(Fig.~\ref{fig:fig2}a). We systematically study how the width of the
distribution of interactions affects the solution of the problem (see Figs. \ref{fig:fig2}b-f
and SI Section \ref{sec:phase_diagram}, for details). 

We simulate the dynamics of directed and weighted mutualistic systems
spanning a large range of parameters across almost two orders of
magnitude in the death rate: $d=0.05$ to $d=4$ (the range of values of
$d$ used in the simulations covers beyond the range of field
measurements which are typically within $d=0.1-0.3$
\cite{thebault,holland}). We use different uniform distributions
$P(\gamma_{ij})$ parametrized by the width $\Delta$ spanning from
$\Delta=0$ (corresponding to a unweighted system where all
interactions are equal, $\gamma_{ij}=\gamma$) to a system with the
widest possible distribution of $\gamma_{ij}$ corresponding to the
maximum width $\Delta_{\rm max}$ allowed by the
condition $\Delta < \gamma$, which is necessary for a mutualistic
system where all interactions are positive, $\gamma_{ij}>0$. The
self-limitation parameter $s$ can be absorbed into the definition of
$d$ and $\gamma_{ij}$ by dividing both parameters, therefore, without
loss of generality, we fix $s=1$ in the simulations, this has the only
effect of changing the unit of measure of the average density by a
factor $1/s$.

A feasible, stable and non-zero solution is found for
$K_{\gamma}<K_{\gamma_c}$.  By feasible, it is understood that the
densities $x_i^*$ must be non-negative, i. e. $x_i\geq0$, for all
species $i$ \cite{sheffer1,sheffer2,thebault, bastolla}.  A necessary
(but not sufficient) condition for the survival of species, and thus
for the existence of a feasible nontrivial fixed point
$\vec{x}^*\neq\vec{0}$, is that $d<\gamma$. This means that the
maximal mutualistic benefit supply of growth factors provided by the
interacting species, corresponding to the interaction strength
$\gamma$ of the nonlinear interaction term, must be larger than the
death rate $d$.

A comparison performed in Figs. \ref{fig:fig2}b-f
(Fig. \ref{fig:fig2}g shows the case $\Delta=0$) between the numerical
solutions for a wide range of parameters and the logic approximation
(black curve) shows a good agreement between the theoretical and
numerical solution. Figure \ref{fig:fig2}h plots the
predicted tipping point for this network $k_{\rm core}^{\rm max}=4$
compared with the numerical solution and shows that the logic
approximation agrees well with the numerical solution for realistic
values of death rates $d \in [0.1-0.3]$\cite{thebault,holland} (SI
Section \ref{sec:limits} elaborates on these results). We estimate that
real ecosystems can be approximated by $\Delta=0$, i.e., the
variability in the interaction term does not significantly affect the
tipping point location for realistic values of the death rate. Second,
the $n=1$ interaction term can be replaced by the logic approximation.
These two approximations allow to obtain the exact solution of the
fixed point equations.

In SI Section \ref{sec:limits} C-D we also consider more realistic distributions
like right-skewed distributions found empirically in Bascompte {\it et
  al.}  \cite{bascompte-linear}.  We integrate numerically
Eqs.~\eqref{eq:model} via a $4^{\rm th}$-order Runge-Kutta
algorithm until the system reaches the fixed point (the simulation procedure
is similar to the one explained in SI Section~\ref{sec:phase_diagram} with $P(\gamma_{ij}) $
empirically measured in \cite{bascompte-linear}).


\clearpage

\begin{figure}[h]
\includegraphics[width=\textwidth]{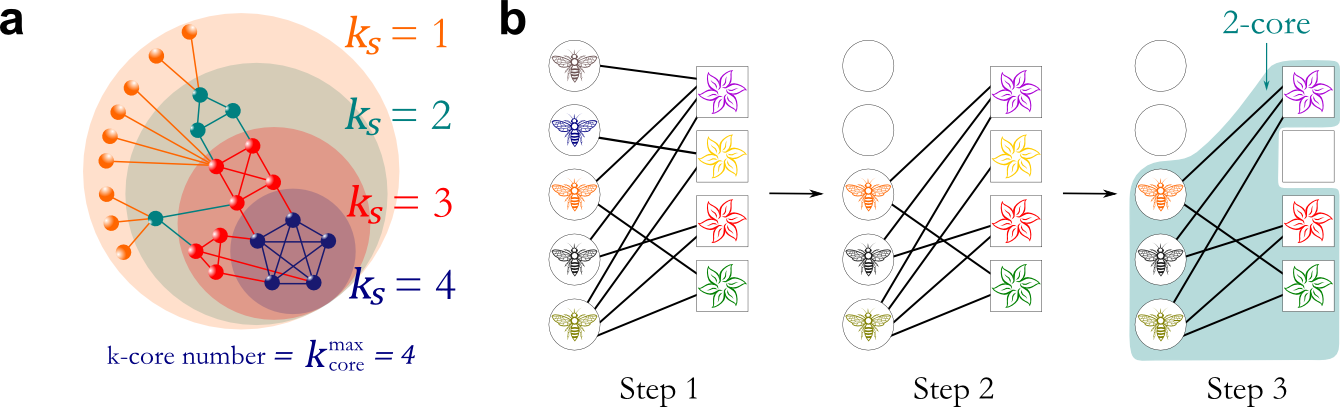} 
\caption{}
\label{fig:fig1} 
\end{figure}

\begin{figure}[h]
\includegraphics[width=\textwidth]{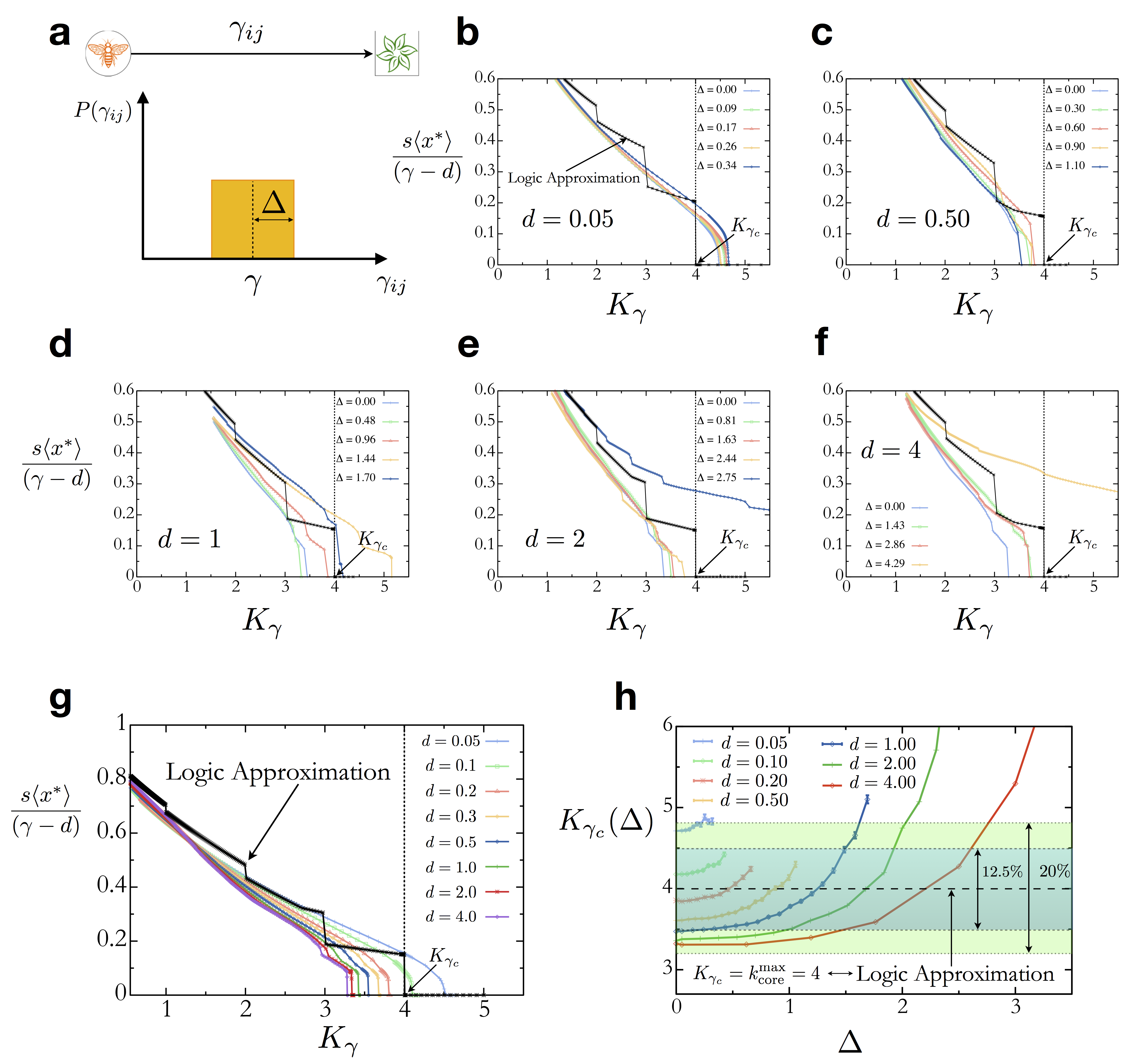} 
\vspace*{-2mm}
\caption{}
\label{fig:fig2} 
\end{figure}

{\bf Fig.~\ref{fig:fig1}}. {\bf k-core structure of a mutualistic network}. 
{\bf a}, Schematic representation of a network as
successive enclosed k-cores, which are the largest subgraphs of the
network where each species is connected at least to $k$ other species .
Species are classified by their k-shell $k_s$, which is the value $k$
of the higher order k-core to which they belong.  The maximum value
$k_{\rm core}^{\rm max}$ attainable by $k_s$ defines the k-core number
of the entire network ($k_{\rm core}^{\rm max}$= $4$ in this case).
{\bf b}, Schematic example of a plant-pollinator mutualistic bipartite
network and the pruning process for extracting the 2-core.  At Step 1
the full network is a 1-core. Then, we remove all species with degree
1, consisting of the two pollinators in the upper left circles (Step
2). These removals produce a new one-degree species, which is the
yellow plant on the right in Step 2. Thus, at Step 3, we remove this plant as
well. The network at Step 3 consists of species of degree 2 or larger,
so the pruning process stops and the result is the 2-core, while the
three removed species constitute the $k_s=1$ shell.

\bigskip

{\bf Fig.~\ref{fig:fig2}}. {\bf Numerical solution to the fixed
    point equations in weighted and directed networks}.  {\bf a},
  Definition of the directed interaction strength $\gamma_{ij}$ from a
  plant $i$ to a pollinator $j$. The interaction strengths
  $\gamma_{ij}$ are i.i.d. random variables drawn from a uniform
  distribution $P(\gamma_{ij})$ with mean $\gamma$ and width $\Delta$.
{\bf b}, Fixed point average density (properly rescaled) $\langle
x^*\rangle=N^{-1}\sum_i x_i^*$ as a function of
$K_{\gamma}=\frac{\alpha s(\gamma+d)}{(\gamma-d)^2}$ (which is
proportional to the inverse average interaction strength $1/ \gamma$,
for small $d$) for the network of a plant-pollinator mutualistic
ecosystem located in the Chilean Andes ~\cite{arroyo} (Net \# 10 in
Supplementary Table~\ref{table:mutualistic_net}), obtained
by solving numerically Eqs.~\eqref{eq:model} using a $4^{\rm
  th}$-order Runge-Kutta algorithm. The death rate is $d=0.05$.  Each
curve is computed using a different sample of interaction strengths
$\{\gamma_{ij}\}$ with a different width $\Delta$ as defined in {\bf a}. For
$\Delta=0$ all $\gamma_{ij}$ are equal. The largest value
$\Delta=0.34$ corresponds to the maximal possible width compatible
with mutualistic interactions, i.e. such that all $\gamma_{ij}$ are
non-negative (details of the numerical simulations in SI Section~\ref{sec:phase_diagram}). We also plot the
analytical solution obtained with the logic approximation (black line).  {\bf c-f},
Same as in {\bf b}, but using a death rate $d=0.5, 1, 2, 4$,
respectively, together with the logic approximation solution.  
{\bf g}, Fixed point average density $\langle x^{*}\rangle$ properly rescaled
as a function of the threshold $K_{\gamma}$ for several values of $d$ and $\Delta=0$.
For comparison, the analytical solution obtained through the logic approximation, 
i.e. solution of Eqs. \eqref{eq:fixpoint_canonical}, or equivalently ~\eqref{eq:y_solution}, 
is plotted as a function of $\langle x^{*}\rangle$ (black line).
The theoretical prediction of the critical value $K_{\gamma_c} = k_{\core}^{\rm max}$ 
is shown in panel {\bf b-g} with a black arrow.
{\bf h},  Critical average interaction strength $K_{\gamma_c}(\Delta)$ as
a function of the width $\Delta$ for different values of $d\in[0.05,
  4.0]$ obtained from Figs. \ref{fig:fig2}b-f.  Each curve ends at a
given value of $\Delta$, which depends on $d$, representing the
maximum admissible width compatible with mutualistic interactions
$\gamma_{ij}\geq0$. Deviations of more than  20\% from the theoretical prediction
given by the logic approximation are found only for large d, i.e. $d>1$, and distribution
width $\Delta >1.5$ (outside the green band in figure).
For values of $d$ of the order of $[0.1-0.3]$, which are the values found in
the literature \cite{thebault,holland}, the theoretical prediction of the logic
approximation are even more accurate and in agreement with the numerical 
simulations of the $n=1$ model within 12.5\%, for any $\Delta$ (blue band in figure).
%
%

\bigskip

\begin{figure}[h]
\includegraphics[width=\textwidth]{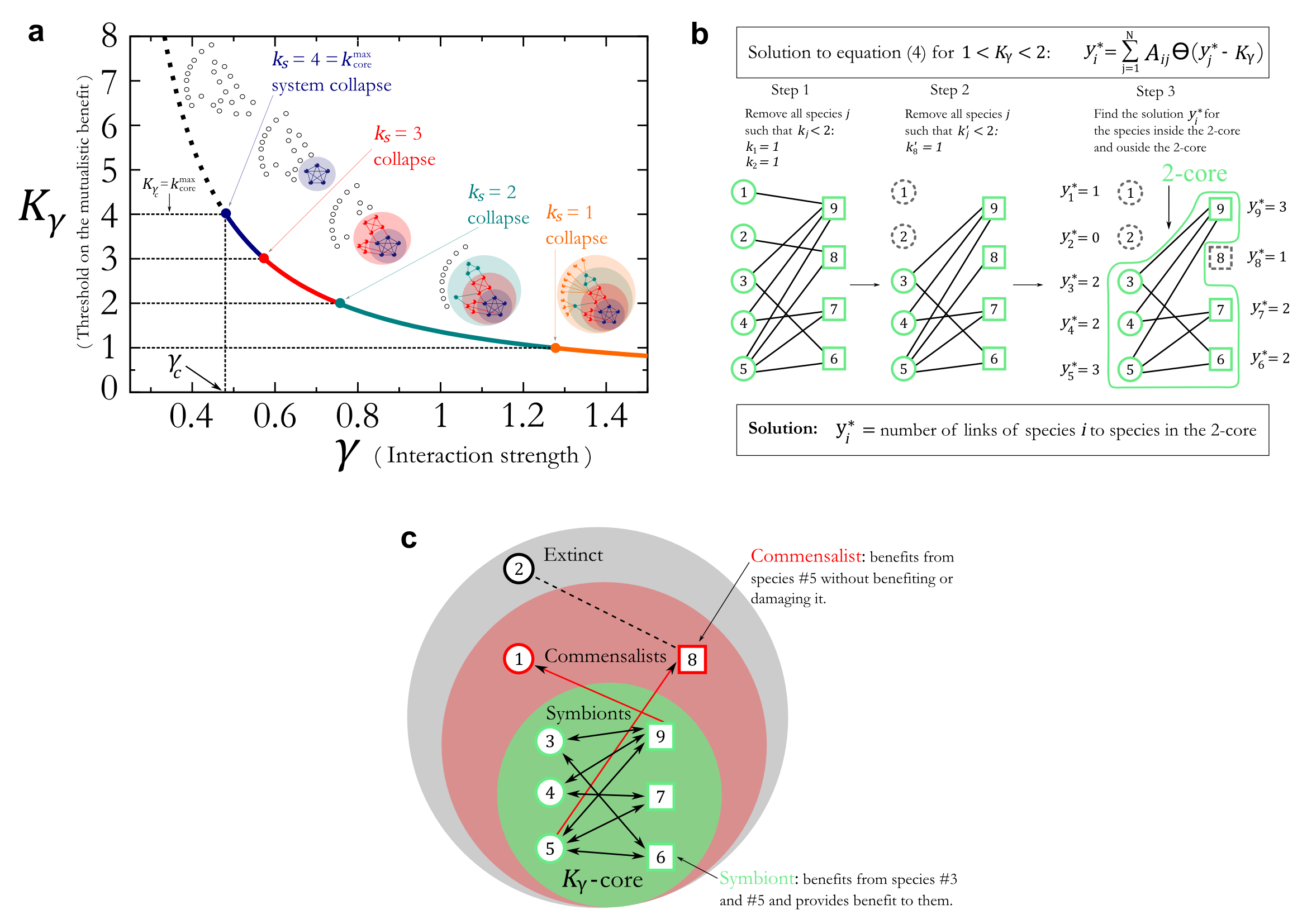} 
\vspace*{-2mm}
\caption{}
\label{fig:fig4} 
\end{figure}

\begin{figure}[h]
\includegraphics[width=\textwidth]{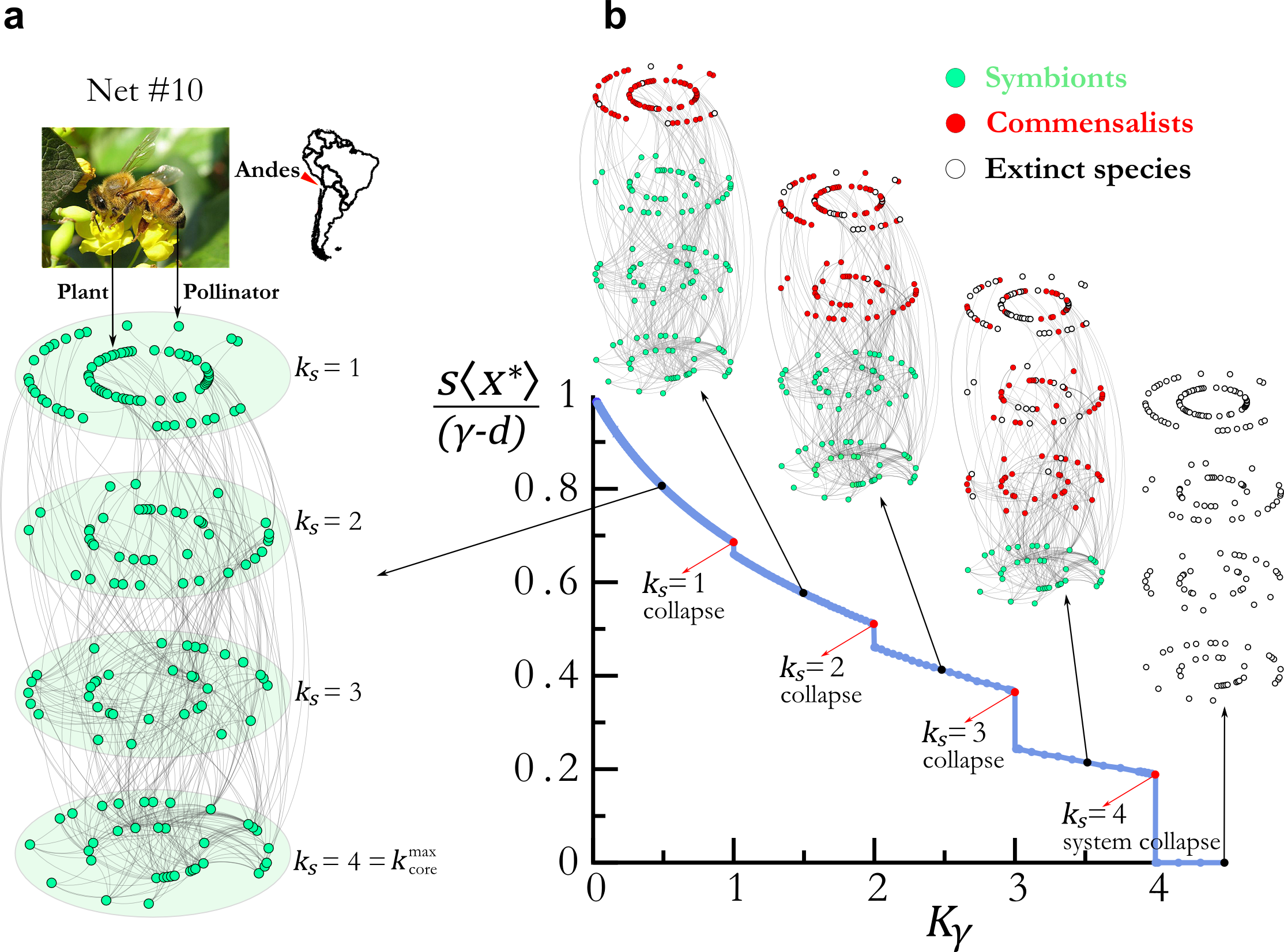} 
\vspace*{-2mm}
\caption{}
\label{fig:fig5} 
\end{figure}

{\bf Fig.~\ref{fig:fig4}}. {\bf Solution scheme for the fixed point
  equations~\eqref{eq:fixpoint_canonical}.} 
{\bf a}, Threshold
$K_{\gamma}$ as a function of the interaction strength $\gamma$ for
the network displayed in {\bf a}. For large $\gamma$ such that
$K_{\gamma}<1$, all species in the network provide their mutualistic 
benefit to the species they interact with.
If $\gamma$ is reduced such that $K_{\gamma}$ is slightly above $1$, 
all the species with $k_s=1$ cannot confer their benefits to the others, 
while the species in the 2-core keep providing their benefit. 
When $\gamma$ is further reduced, so
that $K_{\gamma}$ becomes slightly larger than 2, also the species with 
$k_s=2$ cease to provide their benefit, while the species in the 3-core 
are still able to dispense theirs. 
Further reducing $\gamma$ inhibits the mutualistic benefit from species in 
the 3-shell, too, and, eventually, causes the threshold $K_{\gamma}$ to 
surpass the value of the k-core number of the network 
$K_{\gamma} >k_{\rm core}^{\rm max}=4$,
at which point all the system collapses, since no species can provide 
the mutualistic benefit anymore. This series of collapses results in the 
stair-case shape of species density shown in Fig.~\ref{fig:fig5}b.
{\bf b}, For the sake of explaining
our solution we consider a simple ecosystem network that contains a
2-core and species with interaction strength $K_\gamma$ anywhere
between $1<K_\gamma<2$.  {\it Step 1}: we consider the bipartite
network with all species present.  {\it Step 2}: we remove from the
network
all species $j$ having degree $k_j<K_{\gamma}$, 
since the corresponding variables $y_j^*$ give zero contribution to the 
right hand side of Eqs.~\eqref{eq:fixpoint_canonical}. 
In this case we remove the species $1$ and $2$, since 
$k_1,k_2<K_{\gamma}=2$. 
{\it Step 3}: after these first removals, the species left in the
network have smaller degree $k_j'$, and, we perform a new wave of
removals of species $j'$ if $k_j'<K_{\gamma}=2$. So we remove 
species $8$, since $k_8'<2$. At this point the pruning process 
stops, since the degree of the remaining species is larger than or 
equal to $2$. 
These remaining species $3,4,5,6,7,9$ form the $2$-core of the network. 
Since $y_i^*$ in Eqs.~\eqref{eq:fixpoint_canonical} measures the
number of links to this remaining $2$-core, the solution $y_i^*$ for
the species inside the $2$-core is: $y_3^*=2, y_4^*=2, y_5^*=3,
y_6^*=2, y_7^*=2, y_9^*=3$.  {\it Step 4}: once the solution for the
variables inside the $2$-core has been found, we can add back the
removed species and determine the full fixed point solution. In this
case we add back the species $1,2,8$.  To this end, it is sufficient
to notice that, even for the species placed outside the $2$-core,
$y_i^*$ in Eqs.~\eqref{eq:fixpoint_canonical} still measures the
number of links to species inside the $2$-core.  Therefore, since
species $1$ and $8$ are connected to exactly one species in the
$2$-core, we find $y_1^*=1$ and $y_8^*=1$.  Contrary to species inside
the $2$-core, species $1$ and $8$ have no influence in the system,
meaning that their removal does not change the value of any other
variable.  {\bf c}, In ecological terms, species 1 and 8 are
commensalists, as shown schematically here, as opposed to the true
symbionts living in the $2$-core, because they receive a benefit from
the species in the $2$-core but provide no benefit back. Lastly, for
species $2$, we find $y_2^*=0$, since this species has no links to
species in the $2$-core. Hence it represents an extinct species.  This
exact solution is corroborated numerically in Fig.~\ref{fig:fig5}b.

\bigskip

{\bf Fig.~\ref{fig:fig5}}. {\bf Collapse of a plant-pollinator mutualistic
  network and the tipping line of mutualistic ecosystem.}  {\bf a}, A
bipartite mutualistic network of a plant-pollinator ecosystem located
in the Chilean Andes~\cite{arroyo} (Net \# 10 in Supplementary
Table~\ref{table:mutualistic_net}).  The network is formed
by 4 pairs of concentric rings.  Each pair of rings contains species
with the same k-shell $k_s$, ranging from 1 to 4 (analogous to
Figs.~\ref{fig:fig1}a and \ref{fig:fig4}a). The innermost core is at
$k_{\rm core}^{\rm max}=4$.  Species in the inner rings of each
k-shell represent the plants, and species in the outer rings represent
the pollinators.  {\bf b}, Fixed point average density (properly
rescaled) $\langle x^*\rangle=N^{-1}\sum_i x_i^*$ as a function of the
threshold $K_{\gamma}$, Eqs. (\ref{eq:fixpoint_canonical}), for
the mutualistic network in {\bf a}, obtained by numerical integration
(see SI Section ~\ref{sec:phase_diagram}).  For
$K_{\gamma}<1$, all species are extant and provide their mutualistic
benefit to the species they are linked to in the interaction network
(extant species, green solid symbols).  When $K_{\gamma}$ is above 1,
the species in the outer k-shell $k_s=1$ cannot provide their benefit
anymore, since $K_{\gamma} > k_s$.  However, a species with $k_s=1$
can still benefit from species in the higher shells $k_s>1$, and if it
benefits from at least one of them, it is still extant (red species),
otherwise it is extinct (open circles).  The red species are termed
commensalists because they receive the benefit from other species, but
they do not provide any benefit back. Increasing the threshold further
causes more extant species to turn into commensalists or to go extinct
whenever $K_{\gamma}$ raises above integer values of the successive
k-shell. Finally, when $K_{\gamma}$ becomes larger than $k_{\rm
  core}^{\rm max}=4$, there are no species that can provide a
mutualistic benefit, and the whole system suddenly collapses.

\begin{figure}[h]
\includegraphics[width=\textwidth]{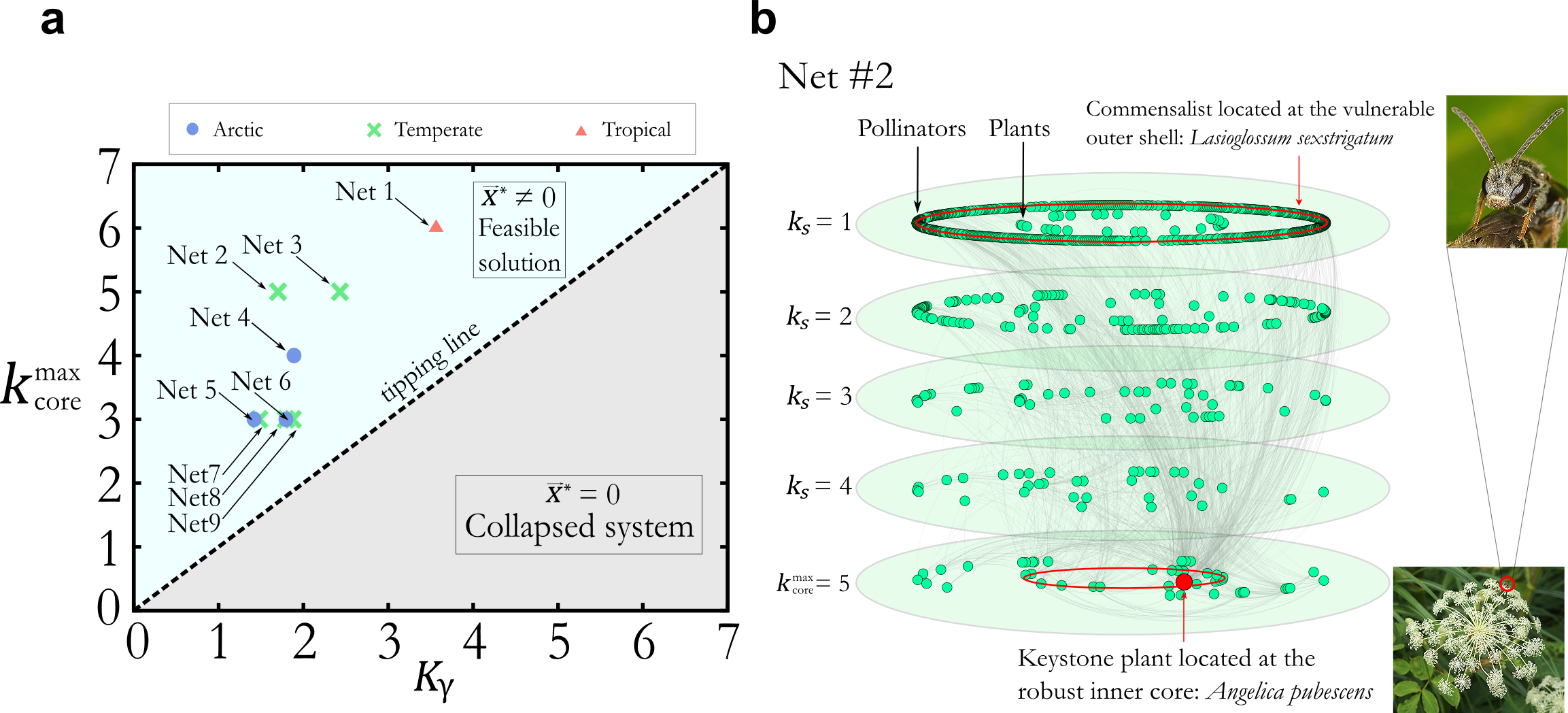} 
\vspace*{-2mm}
\caption{}
\label{fig:fig5cd} 
\end{figure}

{\bf Fig.~\ref{fig:fig5cd}}. {\bf Phase diagram of ecosystem stability}.
{\bf a}, Predicted phase diagram for k-core number $k_{\rm core}^{\rm
  max}$ versus $K_{\gamma}$ for nine empirical mutualistic networks
(Net \#1-9 in Supplementary Table~\ref{table:mutualistic_net})
corresponding to ecosystems at different latitudes: Arctic (blue
point), Temperate (green cross) and Tropical (red triangle).  All the
networks lie in the stable feasible region predicted by Eq.
(\ref{eq:stable}), $k_{\rm core}^{\rm max}> K_{\gamma}$, i.e., above
the tipping line defined by $k_{\rm core}^{\rm max}= K_{\gamma_c}$.
{\bf d}, Shown is the network structure for the ecosystem Net \#2 of
plant-pollinator in Japan from Ref.\cite{kato}. Species are arranged in the
same way as in {\bf a}, i.e., ordered by increasing k-shell number
$k_s$ from top to bottom, with plants in the inner circles and
pollinators in the outer circles. From this graphical representation
an interesting structure emerges.  Many pollinators in the outer shell
$k_s=1$ interact with a single keystone plant species, the {\it
  Angelica pubescens}, located in the innermost core of the network
and therefore is quite stable to external changes, since the inner
core is the most stable core in the ecosystem.  On the contrary, there
are quite fewer plant species in the outer shell ($k_s=1$) interacting
to just a single pollinator in the inner core $k_{\rm core}^{\rm
  max}$).  Plants tend to populate the more robust inner k-shells
while pollinators concentrate more in the low k-shells (i.e. the upper
levels).  This result hints that plants are more needful for the
survival of many pollinators than viceversa, a conclusion stemming
directly from the k-core organization of the ecological network.

\bigskip

\clearpage

\clearpage

\clearpage

\setcounter{section}{0}
\setcounter{figure}{0}

\renewcommand{\figurename}{Supplementary Figure}
\renewcommand{\tablename}{Supplementary Table}

\centerline{ \bf \large Supplementary Information for:}

\centerline{\bf The k-core as a predictor of structural collapse in mutualistic ecosystems}

\medskip

\centerline{Flaviano Morone, Gino Del Ferraro, and Hern\'an A. Makse}

\tableofcontents

\clearpage

\section{Definition of k-core, k-shell and k-core number $k_{\rm core}^{\rm max}$ }
\label{sec:kcore}

The k-core of the network is topologically defined as the maximal
subgraph, not necessarily globally connected, consisting of nodes
having degree at least $k$~\cite{seidman, dorogotsev}. This subgraph 
is unique and can be extracted by iteratively pruning nodes with degree 
less than $k$. By definition, the k-core contains the higher order k+1-core, 
so the 1-core contains the 2-core, the 2-core contains the 3-core, and so on. 
Each k-core is composed by the nodes at the periphery called k-shell and
labeled $k_s$, and the remaining k+1-core.  The periphery of the
k-core is defined as the subgraph induced by nodes and links in the
k-core and not in the k+1-core.  See Figs.~\ref{fig:fig1}a and
\ref{fig:fig1}b for examples of how to calculate the k-cores.

In particular, the 1-shell is a forest, i.e., a collection of trees.
The value $k_{\rm core}^{\rm max}$ of the largest order k-core, which
coincides with the largest value of the k-shell index $k_s$, is called
the k-core number of the network and it corresponds to the innermost
core of the network.  It is a topological invariant of the network,
meaning that it does not depend on how the nodes are labeled or the
network portrayed, i.e., it is invariant under homeomorphisms.
Interestingly, the k-core number is also related to the chromatic
number of the network $\chi$ (defined as the minimum number of colors
to color the nodes so that no neighboring nodes have the same color),
in that $k_{\rm core}^{\rm max}$ provides a bound for $\chi$, i.e.,
$\chi\leq 1+k_{\rm core}^{\rm max}$ \cite{szekeres}. In particular a network is
$\chi$-colorable if it does not have a $\chi$-core (but the converse is
not always true).

\section{Gene regulatory networks and neural networks}
\label{gene}

The specific form of the coupling term in the mutualistic system
defined by Eqs.~\eqref{eq:model} raises the question of what are
the main ingredients necessary for the importance of the k-core for
the tipping point. Thus, it is important to understand how the
specific form of the coupling term in Eqs.~\eqref{eq:model}
affects the main conclusion that the k-core determines the tipping
point of the system.  The model of Eqs. (\ref{eq:model}) is widely
used in ecology \cite{holland,thebault,may,bastolla,holland3} to
describe mutualistic interactions between species and was put forward
in \cite{bastolla}, and then used subsequently by others to study the
stability of ecosystems \cite{thebault}.  The crucial ingredient of
the model is the particular analytic form of the coupling function of
the form $ x_i x_j/(\alpha+\sum_k A_{ik} x_k)$.  We find that the
relevance of the k-core to predict the tipping point is more general
than this particular interaction term. The analytical results are
still valid as long as the interaction term saturates at large values.

For instance, in this Supplementary Information Section \ref{gene2}, we show
that the collapse is predicted by the k-core for a system interacting
via a simpler Hill function coupling of the form $x_j^n/(\alpha^n+
x_j^n)$, which describes expression levels of gene products in
transcriptional networks and in enzymatic reactions
\cite{alon,alon1,alon3,kauffman,kauffman2,gao}. Likewise, we show in
this SI Section \ref{neural}, that other types of sigmoidal
interactions that model the coupling of firing neurons in neural
networks \cite{amit,sompolinsky,gao}: $[1+\tanh(n(x_j-\alpha))]$,
where $\alpha$ is the firing threshold, and $n$ describes the slope of
the sigmoid function, also retains the same dependence, in general
terms, of the tipping point on the k-core as the model of
Eqs.~\eqref{eq:model}.

Below we study these two classes of dynamical systems with different
couplings: gene regulatory networks and neural networks which show the
same type of behaviour as the mutualistic system. All the systems are
described by general response functions that saturate at large values.
It is important to note that below we focus only on the importance of
the shape of the saturating coupling term for the kcore solution. In
particular, we show at the end of this SI Section
\ref{may_stability}, that when one considers the typical linear term of
interaction used in other studies of ecosystems
\cite{may3,allesina,bascompte-linear,coyte}, then a different solution
is found with paradoxical results known as the stability-diversity
paradox \cite{may3}.

Furthermore, in the following examples, we keep the strong condition
that all interactions needs to be positive. Thus, we consider gene
regulatory networks where all genetic interactions are activators and
neural systems of excitatory neurons. Thus, no repressor or inhibitory
interactions are considered in the examples below.  This allows us to
map the dynamical problem to a static problem like k-core percolation,
with the concomitant importance of the giant k-core.  As mentioned in
the main text, systems with excitatory and inhibitory interactions
requires a more general theory beyond percolation, that is presented
elsewhere.


As discussed in the main paper, a functional response widely used in
biology and ecology to model the rate at which $x_i(t)$ changes as a
consequence of the interaction with $x_j(t)$ is the Hill function
$H_n(x_j) = x_j^n/(\alpha^n+x_j^n)$~\cite{alon,holland}.


The parameter $\alpha$ is the activation coefficient, which
defines the minimal density $x_j$ needed to significantly activate the
interaction. The parameter $n$ is the Hill coefficient governing the
steepness of the functional response.

In models of neural networks a popular choice for the functional
response is $G(x_j) =
\frac{1}{2}[1+\tanh(n(x_j-\alpha))]$~\cite{amit}, where $\alpha$
is the firing threshold, and $n$ describes the slope of the sigmoid
function. In particular, for $n\to \infty$, $G(x_j)$ takes only two
discrete values $0$ or $1$, meaning that the neuron is inactive or
firing at the maximum rate and corresponds to the logic approximation
used in Boolean gene networks introduced by Kauffman
\cite{kauffman,kauffman2} and employed in the main text.

In general, we can extend the study of the 
  system defined in the main text to three additional types of
  dynamical systems used in the literature, where the main feature
  is how the rate of change of the activity $x_i(t)$ is modeled 
  by a sigmoidal type of response function~\cite{alon,gao,holland,thebault,amit}:
\begin{equation}
\begin{aligned}
{\rm I}\ \ \dot{x_i}(t)\ &=\ -x_i(d+ s x_i) + \sum_{j=1}^{N}A_{ij}
\gamma_{ij}x_i\frac{x_j^n}{\alpha^n+x_j^n}\hspace{2.8cm} {\rm
  simplified\ mutualistic\ coupling},\\ {\rm
  II}\ \ \dot{x_i}(t)\ &=\ -d x_i + \gamma\sum_{j=1}^{N}  A_{ij}
\frac{x_j^n}{\alpha^n+x_j^n}\hspace{4.7cm} {\rm
  gene\ regulation},\\ {\rm
  III}\ \ \dot{x_i}(t)\ &=\ I - \frac{x_i}{R} +
\frac{J}{2}\sum_{j=1}^{N}A_{ij}
\Big[1+\tanh(n(x_j-\alpha))\Big] \hspace{2cm} {\rm neural\ networks}.
\end{aligned}
\label{eq:models1}
\end{equation}
In the case of neural networks the constants are defined as $I$: the
basal activity, $R$: the inverse of the death rate, and $J$ the
strength of the interactions. Below we elaborate on these models and
show that the k-core determines the tipping point in all of them.

\subsection{Gene regulatory networks}
\label{gene2}

We first study
gene regulatory networks governed by the Michaelis-Menten equation
\cite{alon,gao,karlebach}, where the rate of change of gene expression
$x_i(t)$ can be described by the Hill equation ~\cite{alon,gao}:
\begin{equation}
\begin{aligned}
 \ \dot{x_i}(t)\ &=\ -d x_i\ +\ \gamma_{}\sum_{j=1}^{N}A_{ij}
 \frac{x_j^n}{\alpha^n+x_j^n}\hspace{1cm} {\rm gene\ regulation}\ ,
\end{aligned}
\label{eq:models}
\end{equation}
where $d>0$ is the mortality rate of the genes, $\gamma$ is 
the maximal interaction strength between pair of genes, and the 
activation coefficient $\alpha>0$ defines the minimal expression
activity $x_j$ needed to significantly activate the interaction. The
exponent $n$ of the Hill coefficient governs the steepness of the 
Hill functional response $H_n(x_j, \alpha)$, which is taken as 
$n=2$ or higher~\cite{gao,alon}, thus assuring that the logistic 
approximation is well posed \cite{kauffman,kauffman2}.

To solve the fixed point equations, we thus use the logic
approximation of the Hill function~\cite{alon,amit,kauffman} as in the
main text, $H_n(x_j)\approx\Theta(x_j-\alpha)$, which
is exact for $n\to\infty$.  The step function $\Theta(x)$ equals $1$
if $x>0$ and zero otherwise.  The nonzero fixed point is then:
\begin{equation}
x_i^*\ =\  \frac{\gamma}{d}\sum_{j=1}^{N}A_{ij} \Theta(x_j^*-\alpha)\ , 
\ \ \ \ i = 1, \dots, N\ , 
\label{eq:fixpoint}
\end{equation}
where, for simplicity, we choose uniform dynamical parameters. 
Supplementary Eqs.~\eqref{eq:fixpoint} may be conveniently rewritten using the 
auxiliary variables as in the main text, $y_i^*= x_i^* d/\gamma$, 
and the threshold $K_{\gamma}=(\alpha d)/\gamma$ as
\begin{equation}
y_i^*\ =\  \sum_{j=1}^{N}A_{ij} \Theta(y_j^*- K_{\gamma})\ ,
\ \ \ \ i = 1, \dots, N\ .
\label{eq:fixpoint_canonical2}
\end{equation}
The threshold $K_{\gamma}$ in Supplementary Eqs.~\eqref{eq:fixpoint_canonical2} 
is the bifurcation parameter whose changes produce quantitative and 
qualitative changes of the fixed point solution. 

The solution in this case is obtained in the same way as done in the
main text for the mutualistic ecosystem. First of all, notice that
$y_j^*$ can assume only integer values in the set
$y_j^*\in\{1,\dots,k_j\}$ due to the discrete nature of the step
functions, where $k_j$ is the degree of node $j$.  For a given value
$K_{\gamma}$, we eliminate all the variables $y_j^*$ for which 
$k_j<K_{\gamma}$, since these variables give a vanishing contribution 
to the r.h.s. of Supplementary Eqs.~\eqref{eq:fixpoint_canonical2}, and we only solve 
for the remaining ones.  After this first removal, nodes have smaller degrees
$k_j'$, and if $k_j'<K_{\gamma}$ a new removal occurs until the degrees 
of all the remaining nodes are larger than or equal to $K_{\gamma}$. 
This process is identical to the algorithm for extracting the 
$\lceil K_{\gamma}\rceil$-core of the network~\cite{dorogotsev,gallos}. 
Thus, the nodes left at the end of the pruning process, if there are, form a 
$K_{\gamma}$-core by construction. The solution to the reduced system 
then is obtained by setting all the $\Theta$-functions to $1$, and reads
\begin{equation}
y_i^*\ =\ {\rm numbers\ of\ }i{\rm's\ neighbors\ \in\ }
\lceil K_{\gamma} \rceil-{\rm core}\ , 
\label{eq:y_solution2}
\end{equation}
which is consistent because $\Theta(y_i^*-K_{\gamma})=1$ and we use the
notation: $\mathcal{N}_i(K_{\gamma}) \equiv {\rm
  number\ of\ }i{\rm's\ neighbors\ \in\ } \lceil K_{\gamma} \rceil{\rm-core}$. 
Now we put back in Supplementary Eqs.~\eqref{eq:fixpoint_canonical2} the eliminated
variables.  Since they do not give any contribution to the r.h.s. of
Supplementary Eqs.~\eqref{eq:fixpoint_canonical2}, the solution
Supplementary Eqs.~\eqref{eq:y_solution2} for the in-core variables remains valid
also in the full system.  Moreover, the solution Supplementary
Eqs.~\eqref{eq:y_solution2} is valid also for the out-core variables,
since the nonzero contribution they receive comes from the
in-core variables only. Therefore, the expression of a gene outside the
$\lceil K_{\gamma} \rceil$-core may be non-zero only if it interacts with at
least one of the $\lceil K_{\gamma} \rceil$-core genes.


As in the case of mutualistic networks, the tipping point of collapse of the 
gene regulatory network is obtained at the  critical threshold $K_{\gamma_c}$:
\begin{equation}
k_{\rm core}^{\rm max}
\ =\  K_{\gamma_c}\ \ \to\ \ 
k_{\rm core}^{\rm max}
\ = \  \frac{\alpha d }{\gamma_c},
\label{eq:critical_T2}
\end{equation}
which relates the k-core number $k^{\rm max}_{\rm core}$ of the
regulatory network and the dynamical parameters.  As in mutualistic 
ecosystems, the network structure enters in Supplementary Eq.~\eqref{eq:critical_T2} 
only through the global topological index $k_{\rm core}^{\rm max}$, 
while local details of the network, like the degrees of individual nodes, 
are inessential at the critical point.


\subsection{Neural networks}
\label{neural}

Here we study neural networks governed by the following dynamics
\cite{amit,sompolinsky}:
\begin{equation}
\dot{x_i}(t)\ =\ I - \frac{x_i}{R} +\frac{J}{2}\sum_{j=1}^{N}A_{ij}
\Big[1+\tanh(n(x_j-\alpha))\Big]\ \ \ \ {\rm neural\ networks}\ ,
\end{equation}
where $I$ is the basal activity of the neurons, $R$ is the inverse of 
the death rate, $\alpha$ is the firing threshold, and $J$ is the maximal 
interaction strength between pair of neurons. 
The coefficient $n$ governs the steepness of the sigmoid function, 
analogously to the Hill coefficient $n$ in gene regulatory networks.

To solve the fixed point equations, we use the logistic approximation 
of the response function, $\frac{1}{2}\Big[1+\tanh(n(x_j-\alpha))\Big]
\approx\Theta(x_j-\alpha)$, which is exact in the limit $n\to\infty$.
The fixed point equations then read:
\begin{equation}
x_i^*\ =\ IR +JR\sum_{j=1}^{N}A_{ij}\Theta(x_j^*-\alpha)\ .
\label{eq:fixpoint_neuralnet}
\end{equation}
Supplementary Eqs.~\eqref{eq:fixpoint_neuralnet} can be rewritten using the 
auxiliary variable $y_i^*= (x_i^*-IR)/(JR)$ and the threshold 
$K_J=\alpha/(JR)-I/J$ as 
\begin{equation}
y_i^*\ =\ \sum_{j=1}^NA_{ij}\Theta(y_j^*-K_J)\ ,  
\label{eq:fixpoint_canonical3}
\end{equation}
which is in the same form of Supplementary Eqs.~\eqref{eq:fixpoint_canonical2}. 
Therefore, we can derive the solution of Supplementary Eqs.~\eqref{eq:fixpoint_canonical3} 
by following the same steps after Supplementary Eqs.~\eqref{eq:fixpoint_canonical2}. 
Thus we find:
\begin{equation}
y_i^*\ =\ {\rm numbers\ of\ }i{\rm's\ neighbors\ \in\ }
\lceil K_J \rceil-{\rm core}\ .
\label{eq:y_solution3}
\end{equation}
As in mutualistic and gene regulatory networks, the k-core plays a crucial role 
in the dynamics of neural networks as well. In particular, the tipping point of 
collapse of the neural network is obtained when $K_J$ equals the k-core 
number of the neural network:
\begin{equation}
k_{\rm core}^{\rm max}
\ =\  K_{J_c}\ \ \to\ \ 
k_{\rm core}^{\rm max}
\ = \  \frac{\alpha -IR }{J_cR}\ ,
\label{eq:critical_T3}
\end{equation}
which connects the structure of the neural network, via the 
k-core number $k_{\rm core}^{\rm max}$, to the dynamical parameters, 
in particular the critical interaction strength $J_c$.

\section{Numerical solution in mutualistic weighted and directed networks}
\label{sec:phase_diagram}

%

Interaction strengths between species are, in general, weighted 
and directed, so that $\gamma_{ij}\neq\gamma_{ji}$.  
In this general case it is not possible to find the analytical solution 
to the fixed point equations, so we need to compute this solution 
numerically. 
The fixed point equations of the dynamical system 
Eqs.~\eqref{eq:model} read:
\begin{equation}
x_i^*\ =\  -\frac{d}{s} + \frac{1}{s}\sum_{j=1}^{N}A_{ij}\gamma_{ij} 
\frac{x_j^*}{\alpha+\sum_{j=1}^{N}A_{ij}x_j^*}\ , 
\ \ \ \ i = 1, \dots, N\ , 
\label{eq:fixpoint_general_SI}
\end{equation}
where, as we said, $\gamma_{ij}\neq\gamma_{ji}$ and $A_{ij}\neq A_{ji}$. 
As we explained in the main text, the interactions strengths 
$\gamma_{ij}$ are independent and identically distributed random 
variables drawn from a uniform distribution $P(\gamma_{ij})$ with mean 
$\gamma$ and width $\Delta$: 
\begin{equation}
P(\gamma_{ij})\ =\ \frac{1}{2\Delta}[\Theta(\gamma_{ij}-\gamma+\Delta) - 
\Theta(\gamma_{ij}+\gamma-\Delta)]\ ,
\label{eq:uniform_distribution}
\end{equation}
where $\Delta$ takes value in the interval $\Delta\in[0,\Delta_{\rm max}]$. 
For $\Delta=0$, the interaction strengths $\gamma_{ij}$ are unweighted, 
i.e., $\gamma_{ij}=\gamma$. On the other side, for $\Delta=\Delta_{\rm max}$ 
the interaction strengths are maximally heterogeneous, since $\Delta_{\rm max}$ 
is the maximum admissible width compatible with mutualistic interactions, 
i.e., such that all $\gamma_{ij}$ are non-negative, $\gamma_{ij}\geq 0$. 
Next we explain the procedure to compute the solution to the fixed point 
Supplementary Eqs.~\eqref{eq:fixpoint_general_SI} and how to get the profiles of the 
curves in Figs.~\ref{fig:fig2}b-f.  
\begin{enumerate}
\item For a given $d\in[0.05, 4.0]$, $\gamma\in[0,\infty)$ and
  $\Delta\in[0,\Delta_{\rm max}]$ we draw a sample of
  $\{\gamma_{ij}\}$ from the distribution in
  Supplementary Eq.~\eqref{eq:uniform_distribution} and we assign to each
  directed link the interaction strength $\gamma_{ij}$.
We use $\alpha =1$, and $s=1$. 
  
\item Using the so defined set of $\{\gamma_{ij}\}$ we integrate
  numerically the dynamical Eqs.~\eqref{eq:model} using a $4^{\rm
    th}$-order Runge-Kutta algorithm until all the variables $x_i(t)$
  reach the steady state $x_i(t)=x_i^*$, which is the solution to the
  fixed point Supplementary Eqs.~\eqref{eq:fixpoint_general_SI}.  In the
  Runge-Kutta algorithm we use a time step $\Delta t=0.01$ and we
  iterate the algorithm until the steady state sets in in around
  $10^5$ time steps.  We initialize the densities $x_i(0)>0$ at time
  $t=0$ uniformly at random.
\item We decrease $\gamma$ and repeat step (1) and (2) until the system
  reaches the tipping point of collapse, that is, the fixed point
  $x_i^*=0$ for all $i$.  We denote with $\gamma_c(\Delta)$ the
  critical value of $\gamma$ where the system collapses, and we
  highlight that it depends on $\Delta$.
\end{enumerate}
Thus, following the steps 1-3 we obtain the fixed point solution as a function of 
$\gamma$ for a given $\Delta$, that is $x_i^*=x_i^*(\gamma)$. We measure, in 
the steady state, the average fixed point density 
$\langle x^*\rangle = \frac{1}{N}\sum_{i=1}^Nx_i^*$. In Fig.~\ref{fig:fig2}b we 
show the results of the rescaled average density of species 
$\frac{s\langle x^*\rangle}{\gamma-d}$ as a function of $K_\gamma$ for several 
values of $\Delta$ and $d=0.05$, using the network \#10 in Supplementary 
Table~\ref{table:mutualistic_net}  obtained from 
Ref.~\cite{arroyo}.
Similarly, by changing the value of $d\in[0.05,4.0]$, and repeating the steps 1-3, 
we obtain all other curves depicted in Fig.~\ref{fig:fig2}c-f. Figure \ref{fig:fig2}g shows a similar integration, for several values of $d\in[0.05,4.0]$ with the distribution width $\Delta = 0$.

In Fig.~\ref{fig:fig2}h we show the behaviour of the parameter $K_{\gamma_c}(\Delta)\sim 1/\gamma_c(\Delta)$ as a function of $\Delta$
  for several values of $d\in[0.05, 4.0]$.  The critical parameter $K_{\gamma_c}(\Delta)$ (critical interaction
  strength $\gamma_c(\Delta)$) is defined at the value of $\gamma$ for
  which the average density of species $\langle x^*\rangle$ jumps to
  zero.

\section{Analysis of empirical mutualistic networks}
\label{sec:plant_pollinator} 



Supplementary Table~\ref{table:mutualistic_net} summarizes
the information about the real mutualistic networks used in
Figs.~\ref{fig:fig2}, \ref{fig:fig4}, \ref{fig:fig5}, \ref{fig:fig5cd}.
All the networks analyzed in this work
can be downloaded from the Interaction Web Database at
\url{https://www.nceas.ucsb.edu/interactionweb/}; a nonprofit cooperative
database of published data on species interaction networks hosted by
the National Center for Ecological Analysis and Synthesis, at the
University of California, Santa Barbara, US. This database provides
datasets on species interactions from communities around the
world. Currently available data are for a variety of interaction
types, including plant-pollinator, plant-frugivore, plant-herbivore,
plant-ant mutualists and predator-prey interactions. These data come
from studies in which all species in a particular location, or a
substantial subset, were studied and interactions recorded. The
networks are bipartite webs: species in one group are 
assumed to interact with species in the other group but not with
species in their own group (e.g., plants and pollinators). Each
dataset is defined by an interaction adjacency matrix $A_{ij}$, in
which columns represent one group (e.g., plants) and rows represent
the other group (e.g., pollinators).  We then define the networks of
interacting species via the adjacency matrix $A_{ij}$, which is equal to 
$1$ if species $i$ and $j$ interact, and $0$ otherwise, from where we 
extract the k-core structure of the network.

From this database, we consider only data in the literature for
ecosystems where the full interaction graph $A_{ij}$ has been measured
together with the interaction strength $\gamma$ in order to plot the
networks in the phase diagram of Fig.~\ref{fig:fig5cd}a and test the
feasible-stable condition of Eq.~ \eqref{eq:stable} predicted by the
theory.  The interaction strength $\gamma$ is measured in the field by
counting the frequency of visits of a pollinator to a
plant~\cite{bascompte-linear}, and the actual values can be found in
the Supplementary Information of Ref.~\cite{bascompte-linear}. The values
of the remaining dynamical parameters can be found in
Ref.~\cite{okuyama} and in the Supplementary Table S1 of
Ref.~\cite{thebault}.

The resulting set of real networks is a robust and broad dataset,
comprising of systems located at different latitudes, like Artic,
Temperate and Tropical, different locations from Japan, Australia, USA
to the Chilean Andes and beyond, of relatively large sizes ranging up
to 679 species made of systems of plant-pollinators and plant-seed
dispersers displaying relatively large and robust k-core structures
ranging from $k_{\rm core}^{\rm max}=3$ to 6, as plotted in
Fig.~\ref{fig:fig5cd}a.  We notice that the larger the maximum k-core of
the system, the more robust the system is. That is, for larger $k_{\rm
  core}^{\rm max}$, the system can accommodate a larger decrease in
interaction strength $\gamma$ without collapsing, as seen in
Fig.~\ref{fig:fig5cd}a in the shape of the tipping line in the phase
diagram. These unique datasets that combine network structure and
interaction strengths are ideal to test the predictions of our phase
diagram in Fig.~\ref{fig:fig5cd}a, and supports the main prediction of
the theory regarding the feasible-stable state of ecosystems.




\begin{table}
\begin{tabular}{|c|c|c|c|c|c|c|}
\hline 
Net \# & Network type & Plants & Animals & Latitude & Location  & Ref.\tabularnewline
\hline 
1 & Plant-Seed Disperser  & 31 &  9 &  Tropical &   Papua New Guinea  &\cite{beehl}
\tabularnewline
\hline
2 & Plant-Pollinator  & 91 & 679 &  Temperate &   Japan  &\cite{kato}\tabularnewline
\hline
3 & Plant-Pollinator &   42 & 91 &  Temperate &   Australia  &\cite{ino}\tabularnewline
\hline
4 & Plant-Pollinator &  23 & 118 &  Artic & Sweden  & \cite{elb}\tabularnewline
\hline
5 & Plant-Pollinator & 11  & 18 &  Artic & Canada  & \cite{mos}\tabularnewline
\hline 
6 & Plant-Pollinator  & 14 &  13 & Temperate & Mauritius Island  & \cite{ole}\tabularnewline
\hline 
7 & Plant-Pollinator & 7 &  32  &  Temperate & USA  &\cite{schem} \tabularnewline
\hline 
8 & Plant-Pollinator & 29 &  86  & Artic & Canada & \cite{hocki}\tabularnewline
\hline
9 & Plant-Seed Disperser & 12 & 14  & Temperate & Britain &\cite{soren} \tabularnewline
\hline
10 & Plant-Pollinator &  87 & 99 & Temperate  &  Andes (Chile)  &\cite{arroyo}\tabularnewline
\hline
\end{tabular}
\caption{Details of the 9 mutualistic networks used in the phase
  diagram of Fig.~\ref{fig:fig5cd}a (\#1-9) and the network \# 10 used in
  Fig.~\ref{fig:fig2} and Fig.~\ref{fig:fig5}b.}
\label{table:mutualistic_net}
\end{table}

\section{Derivation of the fixed point solution (6)}
\label{sec:undirected_net}

In this section we show how to derive Eqs.~\eqref{eq:fixpoint_canonical} 
from the nonlinear Eqs.~\eqref{eq:model}, which in turns leads to the 
solution Eqs.~\eqref{eq:y_solution} in terms of the k-core of the network.

From Eqs.~\eqref{eq:model}, there is a trivial fixed point
$x_i^*=0$ for all $i$. This corresponds to the extinction of all
species. The nontrivial fixed point $x_i^* \neq 0$ which corresponds
to the extant species satisfying Eqs. (\ref{eq:y_solution}) is
obtained as follows.  
The equation of the non-trivial fixed point for the dynamical system in 
Eqs.~\eqref{eq:model} with $\gamma_{ij}=\gamma$ reads:
\begin{equation}
x_i^*\ =\  -\frac{d}{s} + \frac{\gamma}{s}\sum_{j=1}^{N}A_{ij} 
\frac{x_j^*}{\alpha+\sum_{j=1}^{N}A_{ij}x_j^*}\ , 
\ \ \ \ i = 1, \dots, N\ . 
\label{eq:fixpoint_SI}
\end{equation}
The trick to find the solution is to turn this set of equations into a
form that can be casted in terms of the Hill function. Then we set:
\begin{equation}
\begin{aligned}
b\ =\ \frac{d}{s} ,\\
c\ =\ \frac{\gamma}{s}\ ,\\
z_i^*\ =\ \sum_{i=1}^{N}A_{ij}x_j^*	\ ,\\
\end{aligned}
\end{equation}
so that Supplementary Eqs.~\eqref{eq:fixpoint_SI} can be rewritten as 
\begin{equation}
z_i^*\ = \frac{\alpha(b +x_i^*)}{(c-b)-x_i^*}\ =\ \sum_{i=1}^{N}A_{ij}x_j^*\ .
\label{eq:fix_point_zed}
\end{equation}
Next, we eliminate $x_j^*$ in favor of $z_j^*$ in the right hand side of 
Supplementary Eqs.~\eqref{eq:fix_point_zed} and we get
\begin{equation}
z_i^*\ =\ \frac{\gamma-d}{s}\sum_{i=1}^{N}A_{ij}\frac{z_j^*-\frac{\alpha d}{\gamma-d}}{\frac{\alpha\gamma}{\gamma-d}+ z_j^*-\frac{\alpha d}{\gamma-d}}\ .
\end{equation}
Finally, we set
\begin{equation}
y_i^*\ = \ z_i^*\frac{s}{\gamma - d}\ , 
\label{eq:y_to_z}
\end{equation}
and we obtain:
\begin{equation}
y_i^*\ = \ \sum_{j=1}^NA_{ij}\frac{(\gamma-d)^2y_j^*-\alpha ds}{\alpha\gamma s + 
(\gamma-d)^2y_j^*-\alpha ds}\ .
\end{equation}
This equations can be written in terms of the Hill function,
$H_n(x,T)$, which is commonly used to describe interactions of species
from ecosystems to biological systems~\cite{alon, amit,kauffman,kauffman2}:
\begin{equation}
H_n(x,T)=\frac{x^n}{T^n+x^n}\ ,
\label{eq:hill_function}
\end{equation}
where $T=\frac{\alpha\gamma s}{(\gamma-d)^2}$ is the half saturation 
constant and $n$ is the Hill coefficient. Using the Hill function, we obtain:
\begin{equation}
y_i^*\ = \ \sum_{j=1}^NA_{ij}
H_1\Bigg(y_j^*-\frac{\alpha ds}{(\gamma-d)^2},\ \frac{\alpha\gamma s }{(\gamma-d)^2}\Bigg)\ . 
\label{eq:y_fix}
\end{equation}

Supplementary Eqs.~\eqref{eq:y_fix} cannot be solved analytically for general 
networks. To find the analytical solution for this fixed point we use 
the logic approximation of the Hill function, which is widely used 
in theoretical biology~\cite{alon, amit,kauffman,kauffman2}: 
\begin{equation}
H_n(x,
T)\approx\Theta(x-T),
\end{equation}
and becomes exact in the limit $n\to\infty$,
where the step function $\Theta(x) = 1$ if $x>0$ and zero otherwise. 
The fixed point Supplementary Eqs.~\eqref{eq:y_fix}, in the logic approximation, 
can be written as follows:
\begin{equation}
\begin{aligned}
y_i^*\ &=\ \sum_{j=1}^NA_{ij}\Theta(y_j^*-K_{\gamma}) ,\\
K_{\gamma}\ &=\ \frac{\alpha s(\gamma+d)}{(\gamma-d)^2}\ ,
\end{aligned}
\label{eq:fixpoint_canonical_SI}
\end{equation}
which is the Eqs.~\eqref{eq:fixpoint_canonical} presented in the
main text. This set of equations admits an exact solution in the form
of Eqs. (\ref{eq:y_solution}) as explained in the main text.
We note that such a solution is valid for any network structure with an
arbitrary degree distribution (such as Erd\"os-Renyi or scale-free
networks) or any internal structure such as modularity, hierarchical,
nestedness, including locally tree-like and dense networks. Thus, it is 
the general exact solution to the fixed point equations of the ecosystem 
dynamics based solely on the logic approximation of the Hill function, 
which allows one to obtain the analytical solution of the problem in closed 
form for any network structure. Then, using Supplementary Eqs.~\eqref{eq:fix_point_zed} 
and~\eqref{eq:y_to_z}, we can also write the nonzero fixed point solution 
Eqs.~\eqref{eq:y_solution} in terms of the original species densities 
$x_i^*$ as:
\begin{equation}
x_i^*\ =\ \frac{(\gamma-d)^2\mathcal{N}_i(K_{\gamma})-\alpha d
  s}{s^2\alpha+s(\gamma-d) \mathcal{N}_i(K_{\gamma})}\ , \ \ \ \ i =
1, \dots, N\ .
\label{eq:x_solution}
\end{equation}



\subsection{Example of solution for systems with 2, 3 and 4 species}
\label{sec:small_net}
In this section we solve the fixed point 
Eqs.~\eqref{eq:fixpoint_canonical} for simple mutualistic
ecosystems with 2, 3 and 4 species shown in 
Supplementary Fig.~\ref{fig:solution_small_net} where the algebra is
straightforward.  This is done to illustrate the solution in a simple
system.

{\bf Ecosystem with 2 species}. The fixed point equations for the 2-species 
ecosystem in Supplementary Fig.~\ref{fig:solution_small_net}a read:
\begin{equation}
\begin{aligned}
y_1^*\ &=\ \Theta(y_2^*-K_{\gamma})\ ,\\
y_2^*\ &=\ \Theta(y_1^*-K_{\gamma})\ .
\end{aligned}
\label{eq:2species}
\end{equation}
The system of Supplementary Eqs.~\eqref{eq:2species} is invariant under a permutation
of species 1 and 2, i.e. for $y_1^*\to y_2^*$. Therefore, we look for a 
homogeneous solution $y_1^*= y_2^*\equiv y^*$ to the single fixed point 
equation:
 \begin{equation}
y^*\ =\ \Theta(y^*-K_{\gamma})\ .
\end{equation}
This equation has the following solution:
\begin{equation}
\begin{aligned}
y^*\ &=\ 1\ \ \ \ {\rm if}\ \ 0< K_{\gamma} < 1\ ,\\
y^*\ &=\ 0\ \ \ \ {\rm if}\ \ K_{\gamma} \geq 1\ , 
\end{aligned}
\end{equation}
which can be rewritten using the function $\mathcal{N}(K_\gamma)$ 
introduced in the main text as 
\begin{equation}
y_1^* = y_2^* = y^* =  \mathcal{N}_1(K_\gamma) = \mathcal{N}_2(K_\gamma)\ . 
\end{equation}
Indeed, the 2-species ecosystem shown in Supplementary Fig.~\ref{fig:solution_small_net}a 
consists only of the 1-core, hence $k_{\rm core}^{\rm max}=1$. 
Therefore, when $K_{\gamma} < k_{\rm core}^{\rm max}$, then $y_1^*$   
is equal to the number of links between species 1 and the species in the 1-core 
$\mathcal{N}_1(K_\gamma)$, which in this case equals 1, since there is only one 
species connected to species 1 in the 1-core.  
On the other hand, when $K_{\gamma} > k_{\rm core}^{\rm max}$, the solution is 
$y_1^* = y_2^* = 0$, in agreement with the general result presented in the main text. 
The same reasoning applies to species 2 by swapping the indices $1\to2$. 

{\bf Ecosystem with 3 species}. The fixed point equations for the 3-species 
ecosystem in Supplementary Fig.~\ref{fig:solution_small_net}b read:
\begin{equation}
\begin{aligned}
y_1^*\ &=\ \Theta(y_2^*-K_{\gamma}) +  \Theta(y_3^*-K_{\gamma})\ ,\\
y_2^*\ &=\ \Theta(y_1^*-K_{\gamma})\ ,\\
y_3^*\ &=\ \Theta(y_1^*-K_{\gamma})\ .
\end{aligned}
\label{eq:3species}
\end{equation}
The system of Supplementary Eqs.~\eqref{eq:3species} is invariant under a permutation
of species 2 and 3, i.e. for $y_2^*\to y_3^*$. Therefore, we look for a 
solution $y_1^*\equiv y^*$ and $y_2^*= y_3^*\equiv z^*$ to the following 
reduced system of equations:
 \begin{equation}
\begin{aligned}
y^*\ &=\ 2\Theta(z^*-K_{\gamma})\ , \\
z^*\ &=\ \Theta(y^*-K_{\gamma})\ ,
\end{aligned}
\end{equation}
whose solution is: 
\begin{equation}
\begin{aligned}
y^*\ &=\ 2\ , z^*\ =\ 1\ \ \ \ {\rm if}\ \ 0< K_{\gamma} < 1\ ,\\
y^*\ &=\ 0\ , z^*\ =\ 0\ \ \ {\rm if}\ \ K_{\gamma} \geq 1\ , 
\end{aligned}
\end{equation}
which can be rewritten using the function $\mathcal{N}(K_\gamma)$ as 
\begin{equation}
\begin{aligned}
y_1^*\ =\ y^*\ = \ \mathcal{N}_1(K_\gamma)\ , \\
y_2^*\ =\ y_3^*\ =\ z^*\ = \ \mathcal{N}_2(K_\gamma)\ = \mathcal{N}_3(K_\gamma)\ . 
\end{aligned}
\end{equation}
Indeed, the ecosystem in Supplementary Fig.~\ref{fig:solution_small_net}b also consists of 
just the 1-core, so that  $k_{\rm core}^{\rm max}=1$. 
Then, when $K_{\gamma} < k_{\rm core}^{\rm max}$,  $y_1^*$   
is equal to the number of links between species 1 and the other species in the 1-core, 
that is species 2 and 3, and thus $y_1^*=2$. Similarly, $y_2^*$ equals 1, 
since it is connected only to species 1, and $y_3^*$ also equals 1 since it 
is connected only to species 1. 
Instead, when $K_{\gamma} > k_{\rm core}^{\rm max}$, the system collapses 
into the trivial fixed point $y_1^*=y_2^*=y_3^*=0$, in agreement with the 
general solution derived in the main text. 

\begin{figure}[h]
\includegraphics[width=\textwidth]{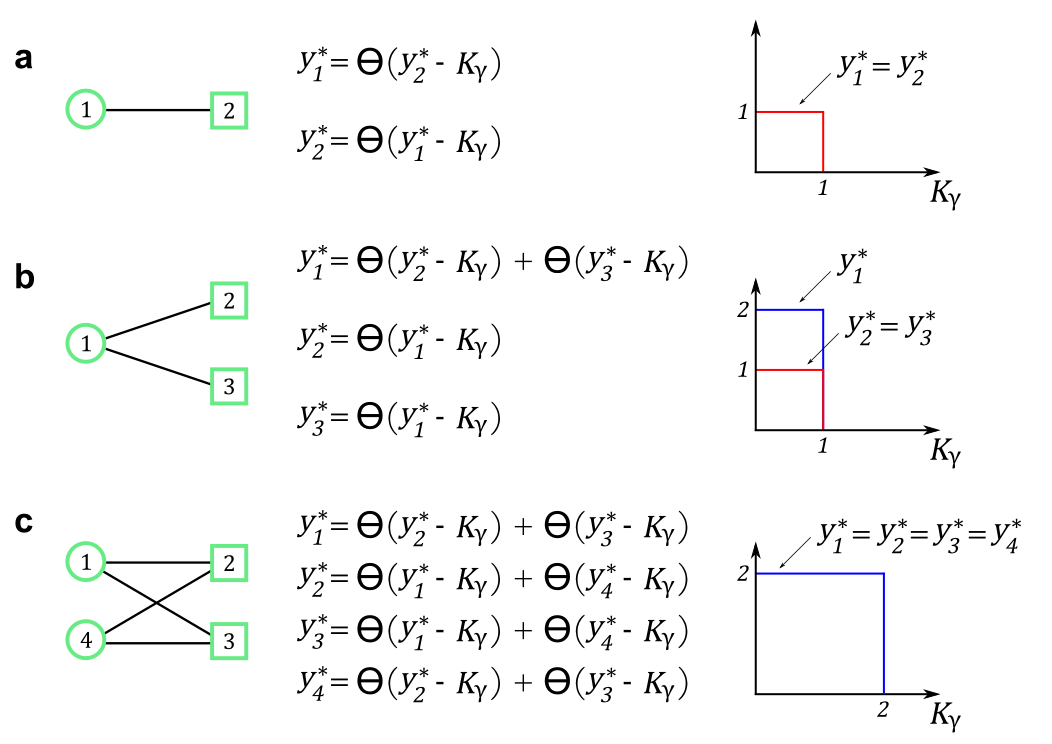} 
\vspace*{-2mm}
\caption{ {\bf Solution to the fixed point equations~\eqref{eq:fixpoint_canonical} 
for small fully connected mutualistic networks.}
{\bf a}, Fixed point equations and the corresponding solution for a fully 
connected mutualistic ecosystem with 2 species. This simple network 
has only the 1-core, so $ k_{\rm core}^{\rm max}=1$. Thus, the system 
collapses when $K_{\gamma} > 1$. The solution for $y_1^*$ and $y_2^*$ 
is given by the number of links connecting species 1 and 2 to the species 
in the $K_{\gamma} $-core, in agreement with the general solution derived 
in the main text.
{\bf b}, Fixed point equations and the corresponding solution for a fully 
connected mutualistic ecosystem with 3 species.  This network has only 
the 1-core, and thus $ k_{\rm core}^{\rm max}=1$. Accordingly, the system 
collapses when $K_{\gamma} > 1$. Also in this case the solution for $y_1^*$, 
$y_2^*$ and $y_3^*$ is given by the number of links connecting species 
1, 2 and 3 to the  species in the $K_{\gamma} $-core.
{\bf c}, Fixed point equations and the corresponding solution for a fully connected 
mutualistic ecosystem with 4 species. This network has only the 2-core, and 
thus $ k_{\rm core}^{\rm max}=2$. Therefore, the system collapses when 
$K_{\gamma} > 2$. Also in this case the solution for $y_1^*$, $y_2^*$, $y_3^*$ 
and $y_4^*$ is given by the number of links connecting species 1, 2, 3 and 4 to 
the species in the $K_{\gamma} $-core. 
}
\label{fig:solution_small_net} 
\end{figure}

{\bf Ecosystem with 4 species}.
The fixed point equations for the 4-species ecosystem in 
Supplementary Fig.~\ref{fig:solution_small_net}c read:
\begin{equation}
\begin{aligned}
y_1^*\ &=\ \Theta(y_2^*-K_{\gamma}) +  \Theta(y_3^*-K_{\gamma})\ ,\\
y_2^*\ &=\ \Theta(y_1^*-K_{\gamma}) +  \Theta(y_4^*-K_{\gamma})\ ,\\
y_3^*\ &=\ \Theta(y_1^*-K_{\gamma}) +  \Theta(y_4^*-K_{\gamma})\ ,\\
y_4^*\ &=\ \Theta(y_2^*-K_{\gamma}) +  \Theta(y_3^*-K_{\gamma})\ .
\end{aligned}
\label{eq:4species}
\end{equation}
The system of Supplementary Eqs.~\eqref{eq:4species} is invariant under permutations
of species 1, 2, 3 and 4. Therefore, we look for a solution 
$y_1^*= y_2^*= y_3^*=y_4^*\equiv y^*$ to the fixed point 
equation:
 \begin{equation}
\begin{aligned}
y^*\ &=\ 2\Theta(y^*-K_{\gamma})\ .
\end{aligned}
\end{equation}
This equation has the following solution:
\begin{equation}
\begin{aligned}
y^*\ &=\ 2\ \ \ \ {\rm if}\ \ 0< K_{\gamma} < 2\ ,\\
y^*\ &=\ 0\ \ \ \ {\rm if}\ \ K_{\gamma} \geq 2\ , 
\end{aligned}
\end{equation}
which can be rewritten using the function $\mathcal{N}(K_\gamma)$ as,
\begin{equation}
y_i^*\ =\ \mathcal{N}_i(K_\gamma)\ =\ 2\  \ \ \ i = 1, 2, 3, 4\ . 
\end{equation}
Indeed, the 4-species ecosystem shown in Supplementary Fig.~\ref{fig:solution_small_net}c
consists only of the 2-core, hence $k_{\rm core}^{\rm max}=2$. 
Therefore, when $K_{\gamma} < k_{\rm core}^{\rm max}$, then $y_i^*$   
is equal to the number of links between species $i$ and the species in the 
$K_{\gamma}$-core, i.e. $\mathcal{N}_i(K_\gamma)$, which in this case equals 2.
Finally, when $K_{\gamma} > k_{\rm core}^{\rm max}$ the solution is 
$y_i^*= 0$ for $i=1, 2, 3, 4$, in agreement with the general solution.

\section{Limits of validity of the approach}
\label{sec:limits}

So far we have studied an oversimplified model of natural ecosystems
which allowed us to reach an exact solution in the limit of the logic
approximation to produce simple predictions on the tipping point. It
is important then to understand the limit of validity of the approach
to determine the conditions under which one might expect the
approximations to give accurate results, and under what conditions the
assumptions are not valid.
  
In what follows we study the limit of validity of the following
approximations as well as perform a comparison with other approaches to
predict the tipping point:

\begin{itemize}

\item Test of the logic approximation in replacing the $n=1$ Hill
  function by the Heaviside (Theta) function

\item Test of theoretical predictions for other types of interaction
  terms used in \cite{dakos}

\item Test of predictions for more realistic cases where the
  interaction strengths $\gamma_{ij}$ are distributed with a
  right-skewed distribution as found empirically in Bascompte {\it et
    al.}  \cite{bascompte-linear}
    
    \item Test of predictions for more realistic cases where the death
  rates and self-limiting parameters are not identical for all species
  (pollinators and plants)
  
\item Test of prediction of collapse over different real webs

\item Comparison with other metrics

\item Test of other conditions that are observed in natural systems
  like plant-plant interactions and other forms of reproductive modes

\item Changes in death rate

\item Other comparisons.
  
\end{itemize}

\subsection{Test of the logic approximation}
\label{logic}

One of the most crucial approximations used to derived the k-core
solution is the use of the logic approximation. Thus, it is important
to understand under which conditions the original $n=1$ Hill function
in Supplementary Eqs. (\ref{eq:model}-\ref{eq:fixpoint_y}) can be replaced by the
Heaviside (Theta) function of the logic approximation.

We solve numerically the fixed point Eqs.
(\ref{eq:fixpoint_canonical}) obtained under the logic approximation
and plot in Fig. \ref{fig:fig2}b-f, as well as in
Fig. \ref{fig:fig2}g, the fixed point average density $\langle
x^*\rangle$ as a function of $K_{\gamma}$, obtained by plugging the
result of Eqs. \eqref{eq:fixpoint_canonical} into the Supplementary
Eqs. \eqref{eq:x_solution} (black line in Figs. \ref{fig:fig2}b-f
and in Fig. \ref{fig:fig2}g) .  In the same figures, we compare this
theoretical prediction with the average density $\langle x^*\rangle$
obtained by numerically integrating Eqs. (\ref{eq:model}) with
$\gamma_{ij}$ sampled from a uniform distribution with different width
$\Delta$ and at different values of the death rate $d$ . All these
numerical calculations are made on the same network of
Ref.~\cite{arroyo}.

Using the simulations of Fig. \ref{fig:fig2} we study how the tipping
point $K_{\gamma_c}$ obtained numerically for a system with $n=1$
deviates from the prediction of the theory for that particular
network, which is $K_{\gamma_c} = k^{\rm max}_{\rm
  core}=4$, for the network used in Fig. \ref{fig:fig2}.
The particular form of the distribution of strength $P(\gamma_{ij})$,
as a uniform distribution with width $\Delta$, allows us to
systematically investigate the logic approximation as a function of
the width as well as other parameters. In the next section we will
repeat the investigation of the validity of the logic approximation
for more realistic distributions of the interactions, such as the
right-skewed distribution found in \cite{bascompte-linear},
and with death rate $d$ and self-limiting parameter $s$ no longer
equal across all species.

Each panel \ref{fig:fig2}b-f in Fig. \ref{fig:fig2} shows the
numerical integration of Eqs. (\ref{eq:model}) for a given value of $d$
as indicated.  Each curve shows the integration for a given $\Delta$
and the comparison to the theoretical prediction using the logic
approximation $n\to\infty$ and the approximation of unique interaction
strength $\gamma$ for all the species. In general, we find that the
logic approximation captures well the $n=1$ system for small enough
death rates $d$ for any $\Delta$, while for large enough death rate
deviations are observed and the logic approximation deviates
substantially from the numerical solution. To quantify this situation,
in Fig.  \ref{fig:fig2}h we plot the numerical tipping point
$K_{\gamma_c}$ for the $n=1$ system as a function of $\Delta$ and for
every system with different $d$. We choose (somehow arbitrary) as a
20\% variation as the limit of validity of the theory. Assuming this
arbitrary cut off, we find that for $d>2$, and for sufficiently large
width of the uniform distribution ($\Delta >1.5$), there are
significant deviations from the $K_{\gamma_c}=4$ logic approximation
prediction (see Fig. \ref{fig:fig2}h, green band). This value marks
the limit of validity of the theory.

The reason why the model does not work for large $d$ can be explained
by inspection of Eqs. (\ref{eq:model}).  Indeed, a condition of
validity of the approach is $d\ll \gamma$. In principle, by definition
the species are expected to interact with another species, as a
minimum, one time in their lifetime which implies $d<\gamma$. Thus,
$d=\gamma$ is the limit of validity of the model. Furthermore, it is
realistic to expect (and this is confirmed by values of $d$ in the
literature, see below) that species interact many times within each other
during their lifetime and, therefore, this constraints
the possible values to $d \ll \gamma$.
Indeed, typical values of $d$ in the literature are in the range $d \in
[0.1-0.3]$, as obtained by Thebault and Fontaine, and Holland et al.
\cite{thebault,holland}. In this range the tipping point of the $n=1$
system, for the largest $\Delta$, 
falls in the range $K_{\gamma_c}\in[3.7, 4.5]$, which is closer to the tipping 
point predicted by the logic approximation $K_{\gamma_c}=4$. 
%
At these values of the death rate $d$ the
deviation of $K_{\gamma_c}$ from the theory reduces to 12.5\% 
(see Fig. \ref{fig:fig2}h, blue band).
We then define this latter the limit of validity of the parameter space following the values
found in other studies \cite{thebault,holland}.  
%

This result suggests that the tipping point of the $n=1$ system in
Eqs.~\eqref{eq:fixpoint_y} can be estimated under the logic
approximation of Eqs. (\ref{eq:fixpoint_canonical}), which, being
analytically tractable, allows us to determine the functional
dependence of the tipping point on the network structure and dynamical
parameters, as we show next.  When the death rate becomes of the order
of $\gamma$, then the mutualistic interactions are of the order of the
death rate and the model and approximations break down.


To disentangle the effects of both approximations, i.e. the uniform
distribution $P(\gamma_{ij})$ and the logic approximation, we plot in
Fig. \ref{fig:fig2}g the numerical simulations for $\Delta=0$ with
different $d$ values and its comparison with the logic approximation
result.  We find that in this case the largest deviation of
$K_{\gamma_c}$ from the theoretical prediction $K_{\gamma_c} = k^{\rm
  max}_{\rm core}=4$ is 17\% and appears for the largest as well as
the smallest value of $d$ ($d=4$ and $d = 0.05$, respectively), both
outside the range of experimental $d$-values found in
\cite{thebault,holland}.  For completeness, in Supplementary Fig. \ref{fig:fig2i} we
show the same results of Fig. \ref{fig:fig2}g by plotting directly
$\langle y^{*} \rangle$ as a function $K_{\gamma}$, obtained by
numerically solving Eqs. \eqref{eq:fixpoint_canonical}.

\begin{figure}[h]
\includegraphics[width=\textwidth]{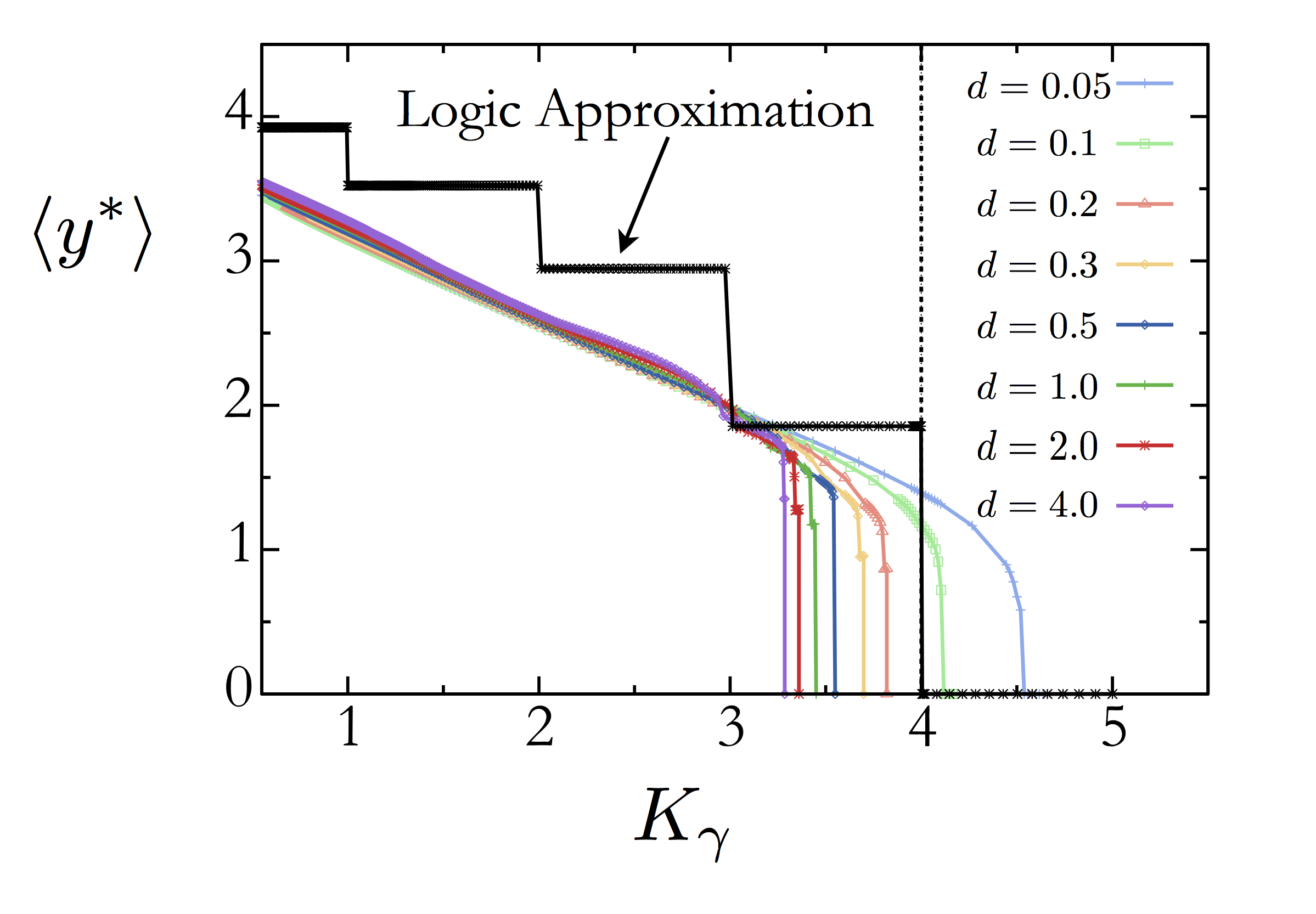} 
\vspace*{-2mm}
\caption{Same results as in Fig. \ref{fig:fig2}g, plotted w.r.t.
  the averaged density $\langle y^{*} \rangle$ obtained from Eqs.\eqref{eq:fixpoint_canonical}.
  Different color lines refer to a numerical integration with a different value
  of the death rate $d$. The black line illustrates the theoretical solution 
  obtained by iteration of Eqs.  \eqref{eq:fixpoint_canonical} till its fixed point.
The values of the self-limiting parameter $s$ and the half-saturation constant
$\alpha$ are the same as in Figure \ref{fig:fig2}, i.e. $s = 1$ and $\alpha=1$.}
\label{fig:fig2i} 
\end{figure}

Let us observe that so far we have studied a critical transition driven by
  an increase of the parameter $K_{\gamma}$ (or equivalently by a
  decrease of the interaction strength $\gamma$). In this case the
  system undergoes an abrupt transition at a tipping point as shown in
  Fig. \ref{fig:fig2}. Figure \ref{fig:fig2}g shows that the logic
  approximation predicts an almost linear decrease of the averaged
  rescaled activity $s<x^*>/(\gamma-d)$ before reaching the tipping
  point at $K_{\gamma_c}=4$. Compared to the logic approximation, the
  numerical solutions in this figure show a slightly different
  singular behaviour of the average density $s<x^*>/(\gamma-d)$ which can still be
  seen as an increase of the magnitude of the first derivative of
  $s<x^*>/(\gamma-d)$ with respect to $K_{\gamma}$ when approaching
  the tipping point (for example, at the parameter $d$ equals 0.05 and
  0.1). 
  
  Yet, when the average density at the fixed point shown in
  Fig. \ref{fig:fig2}g is plotted in terms of the variable $\langle
  y^{*}\rangle$ (Supplementary Fig.  \ref{fig:fig2i}), the numerical solution
  obtained by integrating Eqs. \eqref{eq:model} shows a similar sharp
  transition at the tipping point to the logic approximation (black
  line). Between each shell, the activity measured by the logic
  approximation stays constant and suddenly drops when a given shell
  goes extinct, i.e. $K_{\gamma} = 1, 2, 3, ...$ (see black line in
  Supplementary Fig.  \ref{fig:fig2i}).  The last jump towards a $\langle
  y^{*}\rangle = 0$ solution in the logic approximation is due to the
  collapse of the $k_{\rm core}^{\rm max}$ of the system, in this case
  the $k = 4$ core.  Differently, the numerical solution does not
  presents sharp jumps at $K_{\gamma} = 1, 2, 3,$ but shows a
  progressive linear decrease of the solution (colored lines in Supplementary Fig.
  \ref{fig:fig2i}).  The difference of these two behaviors is due to
  diverse solutions obtained for $n=\infty$ and $n=1$. Indeed, in the
  numerical solution obtained for $n=1$ the k-shells do not collapse
  one by one, since the interaction term in the equation of motion is
  not as sharp as the Heaviside (Theta) function used in the logic
  approximation. Nevertheless, the transition at the tipping point
  towards the ecosystem's collapse is depicted similarly by the two
  solutions because at this point, even the simulated system passes
  abruptly from a non-zero solution to a zero or a non-physical
  solution.  This behavior looks similar to the critical behavior
  observed near critical point of the first-order phase transitions
  discussed in Refs. \cite{shlomo,sheffer1,sheffer2,gao} which is an
  evidence for avalanches in the system. Critical behavior is
  crucially important since it can give early warning signals that may
  occur near critical points of first-order phase transitions in a
  wide class of systems. Thus, further refinements of the model should
  include the study of avalanche behavior as warning of the proximity
  of the tipping point \cite{shlomo,sheffer1,sheffer2,gao}.







\subsection{Test of theoretical predictions for other types of interaction
  terms used in \cite{dakos}}

It is also important to understand how the prediction of the k-core
for the collapse of the system is affected by different models used in
the literature. While Eqs. (\ref{eq:model}) have been studied in
the literature \cite{holland,thebault,may,bastolla,holland3},
other authors have considered modified equations
\cite{dakos}:

\begin{equation}
\dot{x_i}(t)\ =\ -d x_i - s x_i^2 + \sum_{j=1}^{N}A_{ij}\gamma_{ij}
\frac{x_ix_j}{\alpha+\sum_{k=1}^{N}\gamma_{ik}A_{ik}x_k}\ , \,\, i\in\{1, \cdot
\cdot \cdot, N\}\ ,
\label{type2}
\end{equation}
which represent a proper ``Type II'' functional response (see for
example \cite{dakos}). Therefore, it is important to know whether
the results holds for this kind of equations as well.

Using the same approximations employed in Eqs. (\ref{eq:model}) applied
to Supplementary Eqs. (\ref{type2}), we find that by a change of variables ($\alpha
\to \alpha/\gamma$, and $\gamma=1$) the condition
Eq. (\ref{eq:critical_T}) for collapse now becomes:

\begin{equation}
K_{\gamma_c}\ =\  k_{\rm core}^{\rm max}\ \ \to\ \ 
\frac{\alpha s}{\gamma_c} \frac{(1+d)}{(1-d)^2}\ =\  k_{\rm core}^{\rm max}\ ,
\label{kgammac2}
\end{equation}
which represents similar dependence $K_{\gamma_c} \approx \alpha s/
\gamma_c$ as Eqs. (\ref{eq:model}) in the limit of $d\ll\gamma_c$ which
is expected experimentally, since the death rate of the species is
always smaller than the frequency of interactions
\cite{thebault,holland}. Supplementary Information Section \ref{gene} discusses other
variants of systems of coupled equations with similar conclusions:
in all cases the collapse is given by the k-core and the condition of
collapse is inversely proportional to the interaction strength in the
limit of small death rate.

\subsection{Test of right-skewed distribution of $\gamma_{ij}$ from Bascompte {\it et al.} \cite{bascompte-linear}}
\label{sec:right-skewed}

In Supplementary Fig. \ref{fig:fig3}a we present the $P(\gamma_{ij})$ distribution
experimentally obtained from the data in \cite{bascompte-linear} which
shows a right-skewed shape for the distribution of interaction
strengths. Since this distribution is found in Nature, it is of
interest to determine whether and how it affects the results of the
theory. As for Fig. \ref{fig:fig2}, we integrate Eqs. \eqref{eq:model}
but this time with the interaction strengths $\gamma_{ij}$ sampled
from the empirical distribution of \cite{bascompte-linear} shown in
Supplementary Fig. \ref{fig:fig3}a. In order to span several values of $K_{\gamma}$
and produce the plots in Supplementary Fig. \ref{fig:fig3}, we change the average
$\gamma$ by changing the minimal value of the distribution weights
accordingly. The underlying network, as for Fig.  \ref{fig:fig2}, is
the network of Ref.~\cite{arroyo}.  Supplementary Fig. \ref{fig:fig3}b shows the
results for the case of death rate $d$ and self-limiting parameter $s$
equal across all species. As comparison, we also plot the result of
the logic approximation (black line). As the figures shows, the
empirical tipping point $K_{\gamma_c}$ deviates at most by 20\% from
the theoretical predicted one, i.e. $K_{\gamma_c} = k^{\rm max}_{\rm
  core}=4$. We observe that, the use of the experimental distribution
of \cite{bascompte-linear} limits the range of parameters which can be
explored. Indeed, for the experimental $P(\gamma_{ij})$ as in
Supplementary Fig. \ref{fig:fig3}a, the maximum values of $\bar\gamma$ is
$\bar\gamma_{\rm max} = 1$ to which correspond a minimum value of
$K_{\bar\gamma_{\rm min}} = \alpha s (1+d)/(1-d)^2$ which, for any
$d$, constraints the minimum range of $K_{\bar\gamma}$ accessible for
the simulations (see the min value of $K_\gamma$ in
Supplementary Fig. \ref{fig:fig3}b). We further note that, although we have
numerically investigated a range of $d$-values $d > 0.3$, when
integrating Eqs. \eqref{eq:model}, with $P(\gamma_{ij})$ as in
Supplementary Fig. \ref{fig:fig3}a, we have not found any non-trivial solution for
this equation for any $d > 0.38$. This is the reason why we do not
show the same values of $d$ considered in Fig. \ref{fig:fig2} ($d =
0.5, 2, 4$).  However, we observe again that the experimental
$d$-values found in \cite{thebault,holland} are $d \in [0.1, 0.3]$ and
are therefore captured by our numerical investigation with a
right-skewed $P(\gamma_{ij})$ shown in Supplementary  Fig. \ref{fig:fig3}.

\subsection{Test of non identical death rates and
  self-limiting parameters}

In this section we present the results obtained by relaxing 
the assumption of equal dead rates and self-limiting
parameters across species. Following the work of Bascompte 
{\it et al.} \cite{bascompte-linear},  we sample these parameters from 
a uniform distribution $P(d_i)$ and $P(s_i)$, respectively (see Supplementary Fig. \ref{fig:fig3}c). The average
values and the widths of these distributions are chosen from \cite{bascompte-linear}.
From this reference, the average $s$ is taken $s=1$, the $d$-value range therein
($d = 1, 2$), instead, does not give any non-trivial solution to Eqs. \eqref{eq:model}, as discussed above. 
Therefore, we chose $d \in [0.05,0.3]$, as we did for the simulations of Supplementary Fig. \ref{fig:fig3}b, which is
within the range of parameters that produces non-trivial numerical solutions of Eqs. \eqref{eq:model}. 
The width of $P(d_i)$  is taken 0, 10, 20, and 30\% of its mean value, whereas 
the width of $P(s_i)$ is taken 0, 10, and 20\% of $\bar s$,
in agreement with \cite{bascompte-linear}.
The underlying network that we use for the numerical simulation is the same as in Fig. \ref{fig:fig2} 
and, as for the simulations  presented in SI Section \ref{sec:right-skewed} and as mentioned above, 
we do not find any non-zero solution of  Eqs. \eqref{eq:model} for values of $d>0.37 $ 
(with non-zero width of the distributions). 
Results are presented in Supplementary Fig. \ref{fig:fig3}d-g for different values of $\bar d$. In each panel,
each color curve refers to a different value for the width of the  $P(d_i)$ and $P(s_i)$ distribution,
as reported in the legend. In each panel we also show the analytical theoretical prediction obtained
with the logic approximation (black line). Overall, we observe that also in the case of $P(\gamma_{ij})$
right-skewed and both death rate and self-limiting parameters not equal across species 
(the most realistic case for the parameters in the model Eqs. \eqref{eq:model}) the theoretical prediction
are in good agreement with the numerical ones, within a certain range of validity. As for the results 
shown in Fig. \ref{fig:fig2}, the largest deviation from the predicted theoretical value 
$K_{\gamma_c}\ =\  k_{\rm core}^{\rm max} = 4$ for this network is about 20\%, being higher only 
for larger values of $d$, as $d = 0.3$.
\begin{figure}[ht!]
\includegraphics[width=0.9\textwidth]{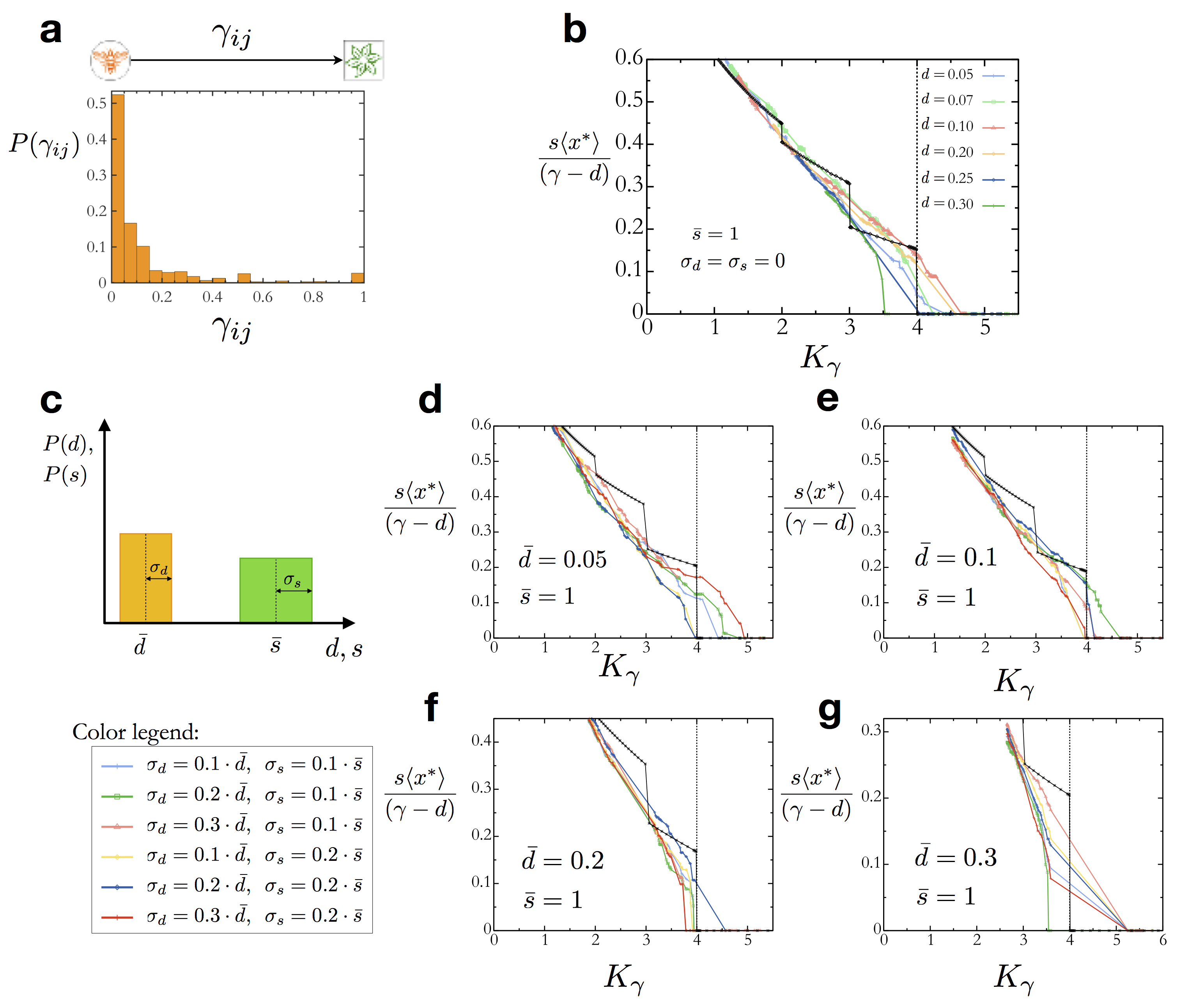} 
\vspace*{-2mm}
\caption{{\bf Test of the theory.} {\bf a,} \small
  Distribution of the interaction strengths for real ecosystems found
  experimentally and reported in \cite{bascompte-linear}. The
  distribution shows a right-skewed shape, which is then used in {\bf
    b,} and {\bf d-g,} to test the theoretical prediction of collapse.
  {\bf b,} Rescaled averaged density $\langle x^{*} \rangle$ as a
  function of $K_{\gamma}$ obtained by numerically integrating
  Eqs. \eqref{eq:model} with $\gamma_{ij}$ sampled from the
  right-skewed distribution of Supplementary Fig. \ref{fig:fig3}a. The death rate
  $d$ and the self-limiting parameter $s$ are taken constant and their
  values are reported in figure. Each curve shows the results of the
  simulation for a different death rate. The black line illustrates
  the theoretical solution obtained with the logic approximation for
  $\bar d = 0.01$, which lies within in the whole range
  considered. {\bf c,} Uniform distribution for the death rates $d_i$
  and the self-limiting parameters $s_i$ which is then used for the
  simulations shown in the following panels. {\bf d-e,} Same numerical
  results as for Supplementary Fig. \ref{fig:fig3}b with death rates and
  self-limiting parameters no longer constant but i.i.d. and sampled
  from the distributions shown in {\bf c}. Mean values are reported in
  figures. Each curve illustrates the results for a different variance
  of the distribution $P(d)$ and $P(s)$, as reported in the
  legend. The black line illustrates the theoretical solution obtained
  through the logic approximation at the $\bar d$-value shown in each
  panel. The underlying network used for the results presented in
  Supplementary Fig. \ref{fig:fig3}b, d-g is the same as the one used in
  Fig. \ref{fig:fig2} (Net \# 10 in Supplementary
  Table~\ref{table:mutualistic_net}).  }
\label{fig:fig3} 
\end{figure}
\subsection{Test of predictions of collapse}
\label{sec:pred_collapse}

To further test our theory, we compare its prediction of the tipping
point for the system collapse with the numerical estimation of the
same tipping point obtained by integrating Eqs. \eqref{eq:model}, with
the right-skewed $P(\gamma_{ij})$ as in Supplementary Fig. \ref{fig:fig3}a and
uniform distribution $P(d_i)$ and $P(s_i)$ for the death rate and
self-limitation parameters, respectively. We use several different
underlying networks of plants-pollinators and plants-seed dispersal
from the Interaction Web Database at
\url{https://www.nceas.ucsb.edu/interactionweb/}.  For each of them we
iterate Eqs. \eqref{eq:model} till the fixed point for different
$K_\gamma$ values until we reach the point of the system's collapse,
i.e. $\langle x^{*} \rangle = 0$, for which $K_\gamma =
K_{\gamma_c}$. Because of the randomness in the interaction strengths,
we repeat the process 30 times for each network and the final value of
$K_{\gamma_c}$ is the average across runs.  We then compare this value
with the theoretical prediction obtained with the logic approximation,
i.e. $K_{\gamma_c} = k_{\rm core}^{\rm max}$.  Results are shown in
Supplementary Fig. \ref{fig:fig6}a where we also report the $R^2$-value for the
linear fit. From this figure we observe that, as predicted by our
theory, the $k_{\rm core}^{\rm max}$ estimates well the point of
collapse for the system ($R^2 = 0.89$) and, therefore, it could be
used as a predictor of the ecosystem's extinction.

\subsection{Comparison with other metrics}
\label{metrics}

We next compare our theoretical solution with other metrics that have
been used to predict the tipping point.  An interesting implication of
the k-core is the fact that k-cores are nested, i.e., high k-cores are
enclosed in low k-cores (Fig. \ref{fig:fig1}a).  According to our
solution, the larger the maximum kcore $k_{\rm core}^{\rm max}$ (i.e.,
the more k-shells in the network) the larger the resilience of the
system against external global shocks that reduce the interaction
strength $\gamma$.  An interesting comparison it is then to study how
$k_{\rm core}^{\rm max}$ and nestedness as defined in
~\cite{bascompte,olesen} correlate with $K_{\gamma_c}$. For each of
the network considered in SI Section \ref{sec:pred_collapse} we compute the
nestedness and plotted it versus the numerical $K_{\gamma_c}$. Results
are shown in Supplementary Fig. \ref{fig:fig6}b and show that the nestedness correlates
weakly with $K_{\gamma_c}$ ($R^2 = 0.01$).

We also note that the level of connectance in the network (i.e., the
average degree) might be important for the collapse: fully connected
networks will collapse abruptly but will have very small $\gamma_c$
(high $K_{\gamma_c}$), while barely connected ones will collapse for
higher values of $\gamma$ (smaller $K_{\gamma}$).  It is then
interesting to study how well $K_{\gamma_c}$ and the connectance
correlate with each other.
 Supplementary Fig. \ref{fig:fig6}c shows significant correlation ($R^2 = 0.56$)
 between $K_{\gamma_c}$ and connectance. Indeed, this is expected
 since the maximum kcore and maximum degree are related by their
 bounds $k_{\rm max} \ge k_{\rm core}^{\rm max}$, and the average
 degree in a given shell is highly correlated with the order of the
 shell, $k_{s}$, for random networks \cite{gallos}.


Furthermore, we observe that there are other quantities that are
related to the $k_{\rm core}^{\rm max}$ of the network and therefore
could be used as proxies of the ecosystem's collapse.  This includes
the spectral radius \cite{staniczenko} and the chromatic number $\chi$
(defined as the smallest number of colors needed to color the vertices
of a graph so that no two adjacent vertices share the same color
\cite{pemmaraju}).  This is because the spectral radius is an exact
upper bound of the $k_{\rm core}^{\rm max}$ (see Ref. \cite{bickle}),
and the chromatic number is a lower bound of $k_{\rm core}^{\rm max}$,
$\chi \leq k_{\rm core}^{\rm max} + 1$ \cite{szekeres}.

Supplementary Fig. \ref{fig:fig6}d
shows the comparison between the spectral radius and the
$K_{\gamma_c}$ for the same networks examined above and illustrates
that also the spectral radius, indeed, correlates well with the
tipping point of the system. In Supplementary Fig. \ref{fig:fig6}e-g, whereas, we
show how this latter metric as well as the others computed above
correlate with the $k_{\rm core}^{\rm max}$. From these results we
observe that, those metrics which are well correlated with the tipping point $K_{\gamma_c}$, i.e. the spectral radius and the connectance (Supplementary Fig. \ref{fig:fig6}c-d), are also well correlated with our theoretical predictor the $k_{\rm core}^{\rm max}$ (Supplementary Fig. \ref{fig:fig6}e-f), as expected by the fact that the $k_{\rm core}^{\rm max}$ is the theoretical predictor for the tipping point of the ecosystem's collapse.

In general, all metrics that are related to $k_{\rm
    core}^{\rm max}$ via mathematical bounds can be approximated
  predictors of the tipping point. However, since these metrics relate
  to the $k_{\rm core}^{\rm max}$ only when the bounds are saturated, they may not
  provide a precise prediction for all type of networks.  While, the
  bounds appear to be saturated in the studied networks, it is not
  guaranteed that other networks will saturate the bounds as well.
  Thus, the $k_{\rm core}^{\rm max}$ remains the only metric that can accurately predict  the tipping point based on first principles for all types of network
  architectures. This fact results from the non-perturbative character
  of our solution which implies that the prediction of the $k_{\rm core}^{\rm max}$ is
  valid independent of the structure of the network.

\subsection{Other limits of validity}

An important condition for the applicability of the k-core solution is
that the system must be mutualistic, that is, all the interactions are
positive $\gamma_{ij}>0$. The interactions can be directed or
undirected, indeed, both cases are solved by different
k-cores. However, the condition of positive interactions is crucial to
introduce the idea of percolation and k-core, and, without this
condition, the k-core percolation cannot be applied to predict the
ecosystem's collapse.  For systems where the interactions can have
positive and negative strengths, like for instance a predator-prey
system, the existence of negative interactions acts as inhibitors in
the system and the concept of the k-core as was derived here cannot 
be applied. 

This case, which is out of the scope of the present
study, leads to other types of fixed points, e.g. limit cycles, and will
be treated in a follow up paper. We anticipate that, while a
topological invariant like the k-core is not anymore relevant in this
case, other invariants arises that allows to connect the structure of
the network to its dynamics. 

There exist other effects that are outside the scope of the present
study, but, nevertheless, they are important and should be
incorporated in future studies. For example, plants are rarely
obligate mutualists and engage in a variety of reproductive modes
which are neglected in our model. The pollinators in the webs
considered here are a subset of those pollinating plants (e.g.,
beetles are rarely included, but contribute quite a bit to
pollination). 
Furthermore, in our model we have considered that plants do not
interact with other plants (nor pollinators with other pollinators).
Future work might incorporate these interesting features. The present
model could be interpreted as how, other things being equal, the k-core
solution may capture the features of the tipping point
\cite{bascompte-linear}. Furthermore, in general, communities of
plants and pollinators are under-sampled in datasets, which might contribute to inaccurate
predictions \cite{bascompte-linear}.

\subsection{Changes in death rate}

We note that Eq.~\eqref{eq:critical_T} is symmetric in $\gamma$
and $d$. As a consequence, the collapse also occurs if the death rate
$d$ increases beyond the critical value determined by
Eq.~\eqref{eq:critical_T}.  Changes in $d$ may be due to
pollution, habitat destruction, genetic isolation or harvesting, which
may be easier to monitor.  However, we observe that the effect of
variations in $d$ on the critical value $K_\gamma$ is much weaker than
the effect caused by variations in $\gamma$.  Indeed, although that
the death rate strongly depends on the life span and can be incredibly
different between plants and animals, both the death rates of plants
and pollinators are much smaller than the relevant scale of the model
which is $\gamma$, in fact $d_P \ll d_A \ll \gamma$, where subscripts
$P$ and $A$ stand for plants and pollinators.  For this reason small
variation of $d$, have very little or none effect on the integer part
of $K_\gamma$ since $d \ll \gamma$ and, therefore, variations in
$\gamma$ are those which dominates the change in $K_\gamma$.

\subsection{Other comparisons}

It would be of interest to further test our
  theoretical prediction based on the k-core using examples of
  collapsed systems lying in the lower region in Fig.~\ref{fig:fig5cd}a,
  i.e. below the tipping line.  This test would require the
  reconstruction of a collapsed mutualistic interaction network via
  fossil records, as has been done for some Cambrian food
  webs~\cite{dunne}.  However, we are not aware of any collapsed
  mutualistic ecosystems whose interaction network has been compiled
  using fossil assemblages. Nonetheless, we notice that the tropical
  network shown in Fig. \ref{fig:fig5cd}a has larger k-core number than
  the temperate and arctic networks, thus being more stable, according
  to our theory.  This result suggests higher resilience in tropical
  networks against extinctions than in temperate or arctic networks, a
  result that has been noticed in experimental studies
  \cite{specialization}.
  In other words, the theory based on the k-core predicts that the
  most vulnerable networks have low k-core number, like the arctic and
  temperate networks in Fig.~\ref{fig:fig5cd}a, and available empirical
  evidence supports this prediction.

\section{Stability of the fixed point solution}
\label{sec:linear_model} 

The fixed point solution derived in the main text is obtained for the
nonlinear system of dynamical Eqs.~\eqref{eq:model}.  Such
nonlinear dynamical equations are characterized by a sigmoid-like
function or Hill function of the interaction term that saturates to a
constant for large densities of the interacting species. 

The stability of the fixed point solution of the nonlinear dynamical 
system~\eqref{eq:model} is controlled by the Jacobian matrix 
\begin{equation}
\mathcal{M}_{ij}(\vec{x}^*)\ =\ \frac{\partial\dot{x}_i}{\partial x_j}\Bigg|_{\vec{x}=\vec{x}^*}\ .
\label{eq:jacobian}
\end{equation}
More precisely, the fixed point solution $\vec{x}^*$ is stable if all 
eigenvalues of $\hat{\mathcal{M}}$ have a negative real part. 
The Jacobian~\eqref{eq:jacobian} for the ecosystem~\eqref{eq:model} 
reads:
\begin{equation}
\mathcal{M}_{ij}(\vec{x}^*)\ =\ -\delta_{ij}\left(d + 2sx_i^*- 
\gamma\frac{\sum_{k=1}^N A_{ik}x_k^*}{\alpha + 
\sum_{k=1}^N A_{ik}x_k^*}\right) + 
\gamma\alpha x_i^*\frac{A_{ij}}{\left(\alpha + 
\sum_{k=1}^N A_{ik}x_k^*\right)^2}\ ,
\label{eq:jacobian_general}
\end{equation}
where we take for simplicity $\gamma_{ij}=\gamma$. 
From Supplementary Eqs.~\eqref{eq:jacobian_general} we see that the trivial fixed 
point $\vec{x}^*=\vec{0}$ is always stable. In fact, in this case, we find 
$\mathcal{M}_{ij}(\vec{0})\ =\ -d\delta_{ij}$, and all eigenvalues equal 
$-d<0$. 

As a consequence, the transition from the fixed point 
$\vec{x}^*=\vec{0}$ to the fixed point $\vec{x}^*\neq\vec{0}$ cannot 
be continuous for any finite value of $d>0$, but must be a discontinuous 
transition. That is, the system must jump from the state $\vec{x}^*\neq\vec{0}$ 
to the state $\vec{x}^*\neq\vec{0}$.

At the nontrivial fixed point, the Jacobian evaluates:
\begin{equation}
\mathcal{M}_{ij}(\vec{x}^*)\ =\ -sx_i^*\ 
\Theta(x_i^*)\left[\delta_{ij} - \frac{\gamma\alpha}{s}\frac{A_{ij}}{\left(\alpha + 
\sum_{k=1}^N A_{ik}x_k^*\right)^2}\right]\ ,
\label{eq:jacobian_nontrivial}
\end{equation}
where the Heaviside (Theta) function $\Theta(x_i^*)$ indicates that
$\mathcal{M}_{ij}$ must be restricted to the extant species, i.e., the
ones such that $x_i^*>0$.  Next, we use the reduced density
$y_i^*=\frac{s}{\gamma-d}\sum_{j=1}^NA_{ij}x_j^*$, and 
Supplementary Eqs.~\eqref{eq:jacobian_nontrivial} as
\begin{equation}
\mathcal{M}_{ij}(\vec{x}^*)\ =\ -sx_i^*\ \Theta(x_i^*)
\left[\delta_{ij} - \frac{\gamma\alpha}{s}\frac{A_{ij}}{\left(\alpha + 
\frac{\gamma-d}{s}y_i^*\right)^2}\right]\ ,
\label{eq:jacobian_y}
\end{equation}
To simplify both notation and interpretation of the subsequent results, 
we notice that, for small $d$, the threshold $K_\gamma$ equals
$K_\gamma=\alpha s/\gamma + O(d)$. Then, taking only the leading 
order in $d$, we can write $\hat{\mathcal{M}}$ as:
\begin{equation}
\mathcal{M}_{ij}(\vec{x}^*)\ =\ -sx_i^*\ \Theta(x_i^*)\left[
\delta_{ij} - A_{ij}
\frac{K_\gamma}{(K_\gamma+y_i^*)^2}\right]\ .
\end{equation}
Using Supplementary Eqs.~\eqref{eq:x_solution} (at the leading order in $d$) 
to express $x_i^*$ in terms of $y_i^*$, we finally obtain $\hat{\mathcal{M}}$ 
as a function of $\vec{y}^*$:
\begin{equation}
\mathcal{M}_{ij}(\vec{y}^*)\ =\ -\gamma\frac{y_i^*}{K_\gamma+y_i^*}\ 
\Theta(y_i^*)
\left[\delta_{ij} - A_{ij}\frac{K_\gamma}{(K_\gamma+y_i^*)^2}\right]\ , 
\end{equation}
which we can rewrite using the Hill function notation as:
\begin{equation}
\mathcal{M}_{ij}(\vec{y}^*)\ =\ -\gamma H_1(y_i^*, K_\gamma)\ 
\Theta(y_i^*)
\left[\delta_{ij} - A_{ij}\frac{H_1(K_\gamma, y_i^*)}{K_\gamma+y_i^*}\right]\ .
\label{eq:jacobian_y_original} 
\end{equation}
At this point we use the logic approximation of the Hill functions. 
Noticing that 
\begin{equation}
H_1(y_i^*, K_\gamma)H_1(K_\gamma, y_i^*)\approx
\Theta(y_i^*- K_\gamma)\Theta(K_\gamma-y_i^*)= 0\ , 
\end{equation} 
Supplementary Eqs.~\eqref{eq:jacobian_y_original} becomes
\begin{equation}
\mathcal{M}_{ij}(\vec{y}^*)\ =\ -\gamma H_1(y_i^*, K_\gamma)\ 
\Theta(y_i^*)\delta_{ij}\ .
\label{eq:final_stab_matrix}
\end{equation}
The eigenvalues of $\hat{\mathcal{M}}(\vec{y}^*)$ can be read directly 
from Supplementary Eqs.~\eqref{eq:final_stab_matrix}, and are given by:
\begin{equation}
\lambda^\mathcal{M}_i = -\gamma\frac{\mathcal{N}_i(K_\gamma)}{K_\gamma+\mathcal{N}_i(K_\gamma)}
\Theta[\mathcal{N}_i(K_\gamma)]\ ,  \ \ \ \ \ i = 1\dots, N\ , 
\label{largest}
\end{equation}
which are Eqs.~\eqref{eq:eigen_stability_matrix} presented in the
main text.  The largest eigenvalue is
\begin{equation}
\lambda^\mathcal{M}_{\rm max}=\max_i \lambda^\mathcal{M}_i \ .
\end{equation} 
The condition for stability of the feasible solution is then 
\begin{equation}
\lambda^\mathcal{M}_{\rm max} < 0 \ ,
\end{equation} 
which guarantees that all other eigenvalues are negative and therefore
the stability of the feasible solution.

According to (\ref{largest}), the largest eigenvalue is attained by
the species with the least number of connections to the
$K_\gamma$-core.  Notice that each eigenvalue is associated with a
single species.  The worst case scenario is a commensalist with just
one link to the $K_\gamma$-core (for instance the commensalists \# 1
and \# 8 depicted in Fig.~\ref{fig:fig4}c) so that the upper bound of the 
largest eigenvalue is
\begin{equation}
\lambda^\mathcal{M}_{\rm max} = -\frac{\gamma}{K_\gamma+1}\ ,
\label{eq:lambda_max}
\end{equation} 
which is always negative. Therefore, the nontrivial feasible solution
$y_i^*$ is always stable, as long as this nonzero solution exists.

On the other hand, when $K_\gamma>k_{\rm core}^{\rm max}$, all
$\mathcal{N}_i$ vanish, i.e. $\mathcal{N}_i(K_\gamma>k_{\rm core}^{\rm
  max})=0$.  Simultaneously, all the eigenvalues become zero,
$\lambda^\mathcal{M}_i = 0$, signaling the simultaneous onset of
collapse and instability of the nonzero fixed point solution.  In
fact, we have discussed after Eqs.~\eqref{eq:y_solution} that the
solution becomes unfeasible and the system must collapse when
$K_\gamma = k_{\rm core}^{\rm max}$.  Thus, we recover from the
stability analysis the tipping point Eq.~\eqref{eq:critical_T},
which we obtained in the main text by requiring the feasibility of the
nontrivial fixed point solution.  That is, we find that the feasible
nontrivial nonzero fixed point solution becomes unstable, 
$\lambda^\mathcal{M}_{\rm max} = 0$, at the same time that it becomes
unfeasible, $K_\gamma = k_{\rm core}^{\rm max}$.




\subsection{Stability analysis of Ref.~\cite{may3}}
\label{may_stability}

Having discussed the stability of our fixed point solution, we discuss
next a method frequently used in the literature to study the stability
of ecosystems modeled as dynamical systems for which the solution to
the fixed point equations is not known~\cite{may3}.  This method
ignores the dependence of the stability matrix $\hat{\mathcal{M}}$
from the fixed point solution, and considers, instead, the alternative
stability matrix $\hat{\mathcal{M'}}$~\cite{may3}:
\begin{equation}
\mathcal{M}'_{ij}\ =\ -\delta_{ij} + \frac{A_{ij}}{K_{\gamma}}\ .
\label{eq:may_matrix}
\end{equation}
Here the adjacency matrix $A_{ij}$ is modeled as a random matrix, giving rise
to a random stability matrix $\mathcal{M}'_{ij}$ and therefore this
method is inspired by Wigner semi-circle law of random matrices.

The stability condition is again that all eigenvalues of $\mathcal{M}'_{ij}$ 
have negative real parts. In this case this stability condition
is expressed by the following:
\begin{equation}
\lambda^A_{\rm max}<K_{\gamma} \,\, \,\,\, \mbox{(condition of stability 
in Ref.~\cite{may3}),}
\label{eq:stability-may}
\end{equation}
where $\lambda^A_{\rm max}$ is the largest eigenvalue of the adjacency 
matrix $\hat{A}$. 
This approximation leads to the so-called diversity-stability
paradox~\cite{may3}, according to which increasing the number of
different mutualistic species will eventually destabilize the ecosystem. This
happens because the largest eigenvalue $\lambda^A_{\rm max}$ of the
matrix $\hat{\mathcal{M'}}$ in Supplementary Eqs.~\eqref{eq:may_matrix}
is a nondecreasing function of the number of different species, and thus the
condition $\lambda^A_{\rm max}<K_{\gamma}$ can be hardly satisfied for
a large diverse ecosystem.  The diversity-stability paradox for
mutualistic ecosystem is, however, 
a by-product 
of the approximative method leading to the stability
matrix~\eqref{eq:may_matrix} which ignores the actual contribution of
the fixed point solution to the stability condition.

Indeed, the exact stability analysis leading to Supplementary Eqs.
(\ref{largest}), by taking into account the dependence of the
stability matrix from the fixed point solution, does not contain the
diversity-stability paradox. On the contrary, it points to the
opposite conclusion that diversity of symbionts increases the
robustness of mutualistic ecosystems.  Specifically, increasing the
number of symbionts who are located in the maximum k-core 
of the network will eventually increase the k-core number 
$k_{\rm core}^{\rm max}$ of their network. As a consequence the 
stability condition Eq. (\ref{eq:stable}): 
$K_\gamma < k_{\rm core}^{\rm max}$ will be easier and easier to satisfy 
as diversity of symbionts increases.
Similarly, mutualistic cooperation stabilizes the system since it
leads to smaller $K_\gamma$ and, again, Eq. (\ref{eq:stable}) is
easier and easier to satisfy as the mutualistic interactions get
stronger.  In conclusion, the stability of mutualistic ecosystem is
primarily controlled by the k-core organization of the underlying
interaction network according to Eq.~\eqref{eq:stable}, where the
diversity-stability paradox disappears and mutualistic interactions are
beneficial for the robustness of the ecosystem.


\begin{figure}[ht!]
\includegraphics[width=\textwidth]{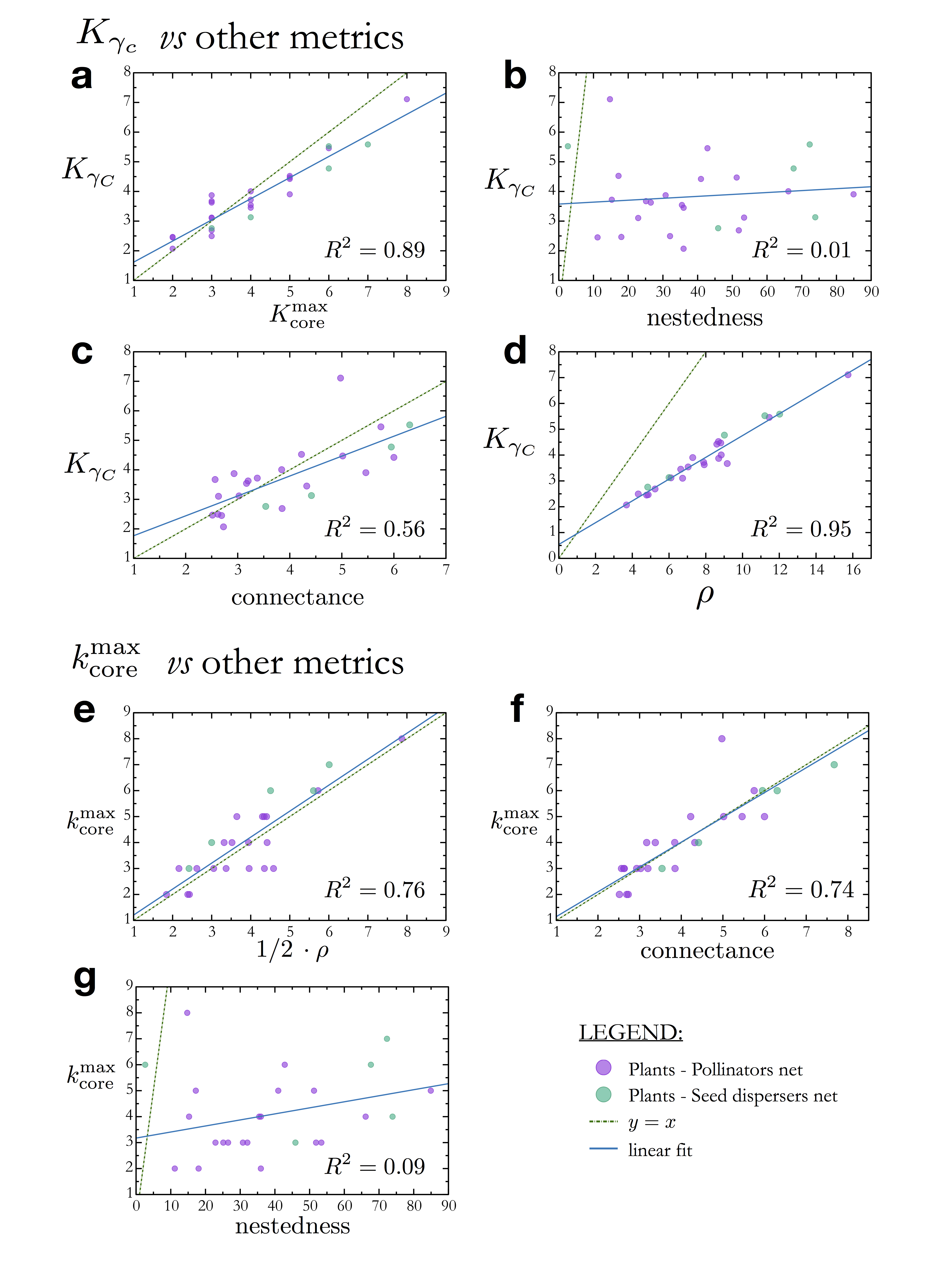}
\caption[entry for the LoF]{Caption in the next page.}
\label{fig:fig6} 
\end{figure}
\clearpage
\noindent
{\small {\bf Supplementary Figure 4: Comparison with other metrics}. {\bf a-d} Comparison
  between the $K_{\gamma_c}$ value at the tipping point of collapse of
  the system and different metrics, for different networks of
  plants-pollinators and plants-seed dispersals from the Interaction
  Web Database at \url{https://www.nceas.ucsb.edu/interactionweb/}. In
  each plot, each point represents the result of the tipping point vs
  a given metric for a specific network. In each panel, $K_{\gamma_c}$
  is compared with: {\bf a,} the $K_{\rm core}^{\rm max}$, {\bf b,}
  the nestedness \cite{bascompte,olesen}, {\bf c,} the connectance
  (defined as the average number of node's connection in the network),
  {\bf d,} and the spectral radius \cite{staniczenko}. The critical
  value $K_{\gamma_c}$ for the tipping point of the system is obtained
  by numerically integrating Eqs. \eqref{eq:model} with
  $P(\gamma_{ij})$ as in Supplementary Fig. \ref{fig:fig3}a till the fixed point and
  by changing the value of $K_{\gamma}$ till one finds the point at
  which all the species go extinct, i.e. $\langle x^{*}\rangle$, (at
  that point $K_{\gamma}$ = $K_{\gamma_c}$). To vary the value of
  $K_{\gamma}$, we change the average $\gamma$ in the $P(\gamma_{ij})$
  by shifting the minimal $\gamma$-value of the distribution weights
  accordingly. {\bf e-g} Comparison between our theoretical predictor
  for the tipping point of collapse of the system, i.e. the $K_{\rm
    core}^{\rm max}$, and other metrics used to characterize
  ecological networks: {\bf e,} the spectral radius, {\bf f,} the
  connectance, {\bf g,} the nestedness. In each panel we plot the line
  $y = x$ and also the line corresponding to a linear fit of the
  data. R-squared values are reported for each figure inside the plot
  frame. Overall, results show that the $k_{\core}^{\rm max}$
  correlates well with $K_{\gamma_c}$ and that those metrics which
  correlate well with the $k_{\core}^{\rm max}$, as the connectance
  and the spectral radius (panel {\bf c} and {\bf d} respectively),
  also correlate with $K_{\gamma_c}$ (see panel {\bf e} and {\bf f}).
  The $k_{\core}^{\rm max}$ and the spectral radius are mathematically
  related, indeed the spectral radius is always an upper bound of the
  $k_{\core}^{\rm max}$ \cite{bickle}.}

\end{document}